\documentclass[twocolumn,showpacs,superscriptaddress,groupedaddress]{revtex4-1}  

\usepackage{graphicx}  
\usepackage{dcolumn}   
\usepackage{bm}        
\usepackage{amssymb}   
\usepackage[export]{adjustbox}

\begin{document}

\widetext

\title{High-order harmonic generation from Rydberg atoms driven by plasmonic-enhanced laser fields}

\author{Y. Tikman*$^1$}
\author{I. Yavuz$^1$}
\author{M. F. Ciappina$^2$}
\author{A. Chac\'on$^3$}
\author{Z. Altun$^1$} 
\author{M. Lewenstein$^{3,4}$}

\affiliation{$^1$Marmara University, Physics Dep. 34722, Ziverbey, Istanbul, Turkey} 
\affiliation{$^2$Max-Planck-Institut f\"ur Quantenoptik, Hans-Kopfermann-Strasse 1, 85748 Garching, Germany}
\affiliation{$^3$ICFO-Institut de Ci\`encies Fot\`oniques, The Barcelona Institute of Science and Technology, Av.~Carl Friedrich Gauss 3, 08860 Castelldefels (Barcelona), Spain}
\affiliation{$^4$ICREA-Instituci\'{o} Catalana de Recerca i Estudis Avan\c{c}ats, Lluis Companys 23, 08010
Barcelona, Spain}



\email{yavuztikman@marun.edu.tr}

\date{\today}

\begin{abstract}
We theoretically investigate high-order harmonic generation (HHG) in Rydberg atoms driven by spatially inhomogeneous laser fields, induced, for instance, by plasmonic enhancement.  It is well known that the laser intensity should to exceed certain threshold in order to generate HHG, when noble gas atoms in their ground state are used as an active medium. One way to enhance the coherent light coming from a conventional laser oscillator is to take 
advantage of the amplification obtained by the so-called surface plasmon polaritons, created when a low intensity laser field is focused onto a metallic nanostructure. The main limitation of this scheme is the low damage threshold of the materials employed in the nanostructures engineering. In this work we propose to use Rydberg atoms, driven by spatially inhomogeneous, plasmonic-enhanced laser fields, for HHG. We exhaustively discuss the behaviour and efficiency of these systems in the generation of coherent harmonic emission. To this aim 
we numerically solve the time-dependent Schr\"{o}dinger equation for an atom with an electron initially in a highly excited $n$-th Rydberg state, located in the vicinity of a metallic nanostructure, where the electric field changes spatially on the scales relevant for the dynamics of the laser-ionized electron.  We first use a one-dimensional model to investigate the phenomena systematically. We then employ a more realistic situation, when the interaction of a plasmonic-enhanced laser field with a three-dimensional Hydrogen atom is modelled. We discuss the scaling of the relevant input parameters with the principal quantum number $n$ of the Rydberg state in question, and demonstrate that harmonic emission could be achieved from Rydberg atoms well below the damage threshold, thus without deteriorating the geometry and properties of the metallic nanostructure. 

\end{abstract}

\maketitle

\section{Introduction}

High-order harmonic generation (HHG) has been thoroughly studied and well-established due to its potential for synthesizing bright isolated xuv pulses, which are the workhorse to understand ultra-fast electron dynamics on the sub-femtosecond and sub-\r{A}ngstrom spatio-temporal scale~\cite{corkum2007attosecond,brabec2000intense}. The dynamics of HHG is transparently described by the so-called three-step model~\cite{corkum1993plasma,lewenstein1994theory}. We could summarize the sequence as follows: (i) a bound electron first tunnels out from the Coulomb barrier, suppressed by the incident laser electric field; (ii) then this laser-ionized electron accelerates in the continuum under the sole influence of the electric field; (iii) finally the electron is driven back towards the parent ion and converts its kinetic energy in energetic and coherent photons upon recombination. The maximum photon energy achievable from the HHG process is determined by the classical cut-off law $\omega_{\mathrm{cut-off}} =  I_p + 3.17 U_p$~\cite{corkum1993plasma,lewenstein1994theory}, where the ponderomotive potential $U_p$ is $\sim I \lambda^2 $ [$I$ is the peak field intensity and $\lambda$ the wavelength of the incident field] and $I_p$ the ionization potential of the target atom of interest. 

The relationship $\omega_{\mathrm{cut-off}}\propto I \lambda^2$ suggests two main routes to obtain high energetic coherent photons, namely (i) to increase the laser intensity $I$ or (ii) to use laser sources with longer wavelengths. Several limitations in both schemes are in order. On one hand, ionization clearly rises for higher laser intensities, which already at the level of single atom response dramatically decreases the probability of electron recombination responsible for emission of photons of high frequency --in effect there is no HHG above the, so-called, saturation intensity $I_{\mathrm{sat}}$ (cf.~\cite{AnneML}).  Moreover, other  instrumental control parameters needed to obtain an appreciable photon flux,  such the phase matching between the atomic emitters (for details see e.g.~\cite{Gaarde,Sanpera1995}), are affected by higher ionization levels.  The other pathway, the utilization of larger wavelengths, suffers from disadvantages as well. For instance, it is demonstrated that the HHG yield scales as $\lambda^{-(5 \sim 6)}$, and, as a consequence, it appears to be challenging to obtain a measurable signal as $\lambda$ increases due to this poor conversion efficiency~\cite{Tate}.

In an ordinary HHG experiment, a supersonic jet of noble gas atoms, all in their ground states, is used as a target ($I_p \sim 15-30$ eV). When near-infrared lasers are employed, the minimum laser intensity to observe the HHG phenomenon lies in the range of $1\times 10^{12 \sim 13}$ W/cm$^2$, which often requires a second stage amplification at the output of a conventional femtosecond laser oscillator (with typical output intensities in the range of $1\times 10^{10 \sim 11}$ W/cm$^2$). Considering the large infrastructure needed for this amplification phase, many alternative schemes have been proposed to amplify the incident field as to permit an efficient strong field-matter interaction~\cite{gohle2005frequency,strickland1985compression}. One that has received great interest recently is the so-called plasmon-enhanced HHG~\cite{schuller2010plasmonics,kim2008high,park2011plasmonic,sivis2013extreme, pfullmann2013bow,park2013generation}. This approach takes advantage of the field enhancement generated when a laser field of moderated intensity is focused onto an array of metallic nanostructures~\cite{hutter2004exploitation}. With this setup it is possible to boost up the input laser power several orders of magnitude~\cite{kim2008high}. Here, a special attention should to be paid to the geometry of the nanostructure, typically a metallic bow-tie of nanometric dimensions with a gap between the apexes (for details see e.g.~\cite{kim2008high}). By fine-tuning the input configurations, such as the spatial metal geometry or the gap distance, the nanostructure elements, acting like point sources, are able to amplify an incident field with \textit{moderate} intensity ($\sim 1\times 10^{11}$ W/cm$^2$) up to the intensities needed for producing XUV emission from atomic gas targets~\cite{kim2008high, park2013generation}.

Due to the confinement of the incoming field into a nano-volume, the resulting plasmonic-enhanced field
presents typically a spatial variation on a nanometer scale. For instance, within a gap of roughly 20 nm, the intensity of the input field is enhanced a few orders of magnitude at the gap center and drastically increases and culminates with a maximum near the metal surface~\cite{kim2008high}. Due to such field gradient and the electron confinement into a small volume, the conventional Keldysh picture of strong-field physics -in which the spatial dependence of the laser field and the influence Coulomb potential is ignored- is incapable to fully explain the underlying physics~\cite{keldysh1965ionization,blaga2009strong,fetic2013high}. As a result, there has been an intense effort to comprehend both the underlying physics at the microscopic level and to numerically simulate the processes at a multi-scale level~\cite{husakou2011theory,ciappina2012high,yavuz2012generation,perez2013beyond,fetic2013high,ciappina2013high,park2013generation,yavuz2013gas,zhang2013control,he2013wavelength,luo2014efficient,ebadi2014interferences}. Elucidating the mechanisms behind plasmon-enhanced HHG are thus essential due its potential technological applications, e.g.~\cite{lupetti2013plasmon}.    

Since the first realization of HHG from plasmon-enhanced fields by bow-tie shaped nano-antennas, a few alternative nano-systems have been explored in order to optimize as well as to modify the phenomenon through the synergy between experiments and theory. Amongst these systems we could cite; coupled ellipsoids~\cite{stebbings2011generation}, tapered nano-cones~\cite{park2011plasmonic, sivis2013generation}, metal nano-particles~\cite{shaaran2013high} or nano-composites~\cite{husakou2014quasi,sakabe2015progress}. However, despite its initial meteoric success and promises, plasmon-enhanced harmonic emission process still suffers in many aspects, such as the melting of the metallic nanostructures caused by the high build-up intensities near the metal surface. Additionally, the low conversion-efficiency due to the low target gas density contributing to the photo-emission makes challenging to increase the signal/noise ratio~\cite{pfullmann2013bow, sivis2013generation}. On the other hand, there are apparent discrepancies between the intensity enhancements predicted through finite-element simulations and those that are necessary for efficient HHG  as estimated by the conventional three-step model~\cite{shaaran2012estimating}. Because of these circumstances, the feasibility to use metallic nanostructures to drive HHG is actively debated in the literature~\cite{raschke2013high,sivis2012nanostructure}. Having these complications in mind, our main purpose is to address an alternative way to obtain efficient HHG, while avoiding the damage on the nanostructure elements caused by the high intensity plasmon-enhanced fields.     

One way to decrease the intensity necessary for efficient HHG is to prepare atoms in their excited states, which are characterized by smaller values of $I_p$. The price one has to pay, however, is that the HHG cutoff decreases in accordance with the classical law $\omega_{\mathrm{cut-off}} =  I_p + 3.17 U_p$, where both $U_p$ and saturation intensity $I_{\mathrm{sat}}$ must necessarily be smaller.  One way to circumvent this obstacle and to even enhance the cutoff is to consider HHG originating from a superposition of an excited and a ground state~\cite{Sanpera1,Sanpera2} or excited vibrational states in molecules \cite{Moreno} -- for the more recent developments of these ideas see~\cite{Milosevic,Milosevic2,Yuan}. Another, easier way is to scale the intensity and the laser wavelength (frequency) in accordance with $I_p$. It is the latter strategy the one we will adopt in this paper. 

Owing to their loosely bound valence electrons, Rydberg atoms are highly sensitive to external influences that can easily cause them to  ionize~\cite{gallagher2005rydberg}. In other words, while high-intensity fields are required to ionize an atom in its ground state $n=1$, the intensity required to ionize a Rydberg atom $n\gg1$ is considerably smaller. Here, the scaling of Rydberg atoms with the principal quantum number $n$ offers insights into their peculiar features \cite{gallagher2005rydberg}. As shown in many studies, the relevant parameters here, e.g.~the ionization potential $I_p$, the radius of the $n$-th Rydberg orbit $r_R$, the electric field strength $E_R$ felt by an electron at the $n$-th Rydberg orbit, and the energy level spacing $\Delta E_R$, the latter leading to even more closely spaced levels as $n$ increases, scale with $n$ as:
\begin{equation}
I_p = n^{-2} \widetilde{I_p},  
\end{equation}
\begin{equation}
r_R = n^{2} \widetilde{r},  
\end{equation}
\begin{equation}
E_R = n^{-4} \widetilde{E_0}  
\end{equation}
and
\begin{equation}
\label{omegascalR}
\Delta E_R = n^{-3} \widetilde{\Delta E_0}, 
\end{equation}
where the tilde quantities correspond to the values for $n=1$.

Thus, if one applies an intense laser field to Rydberg atoms and attempts to study the atomic response as a function of $n$, it is natural to consider the $n$-scaling of the relevant electron dynamics parameters based on the above equations. According to this approach, if ionization potential scales as above $I_p = n^{-2} \widetilde{I_p}$, the laser electric field as we consider higher $n$'s
should scale as
\begin{equation}
E_0 = n^{-4} \widetilde{E_0}  
\end{equation}
and the laser frequency as 
\begin{equation}
\label{omegascal}
\omega_0 = n^{-3} \widetilde{\omega_0}, 
\end{equation}
where the tilde quantities correspond now to characteristic values of the laser field and frequency used for  $n=1$.
In such a situation for instance, the ponderomotive potential $U_p = E_0^2 / 4 \omega_0^2$ scales as $n^{-2}$, and thus the cutoff frequency (or maximum photon energy) as $\omega_{\mathrm{cut-off}}$ and cutoff harmonic order $q_{\mathrm{cut}}=\omega_{\mathrm{cut-off}}/\omega_0$ scales as $n^{-2}$ and $n$, respectively. Interestingly, the Keldysh parameter $\gamma = \sqrt{I_p/2U_p}$, which separates the
tunneling (for $\gamma \ll  1$) and multiphoton (for $\gamma \gg 1$) dynamical regimes, remains unscaled with $n$~\cite{bleda2013high}. Finally, the classical quiver radius of the laser ionized electron $\alpha_0 = E_0 / \omega_0^2$ scales as $n^{2}$, i.e.~like the Rydberg radius $r_R$. 

In this paper, we address and demonstrate the behavior and the usefulness of Rydberg atoms in the plasmonic-enhanced HHG process. Considering the fact that  the tunnel ionization is instrumental in the HHG process~\cite{corkum1993plasma,Gaarde}, the Rydberg atoms could be advantageous for HHG driven by plasmonic-enhanced fields generated by metallic nanostructures, since relatively low intensities are required to detach Rydberg electrons via the tunneling processes. This fact could prevent -potentially- the damage and melting of the employed nanostructures  during the process. 

In the experiments performed in plasmonic-enhanced HHG, the target atoms -mostly in their ground states- are directly injected onto the nanostructure through a gas jet~\cite{kim2008high, park2013generation}.  If we want to drive atoms in an excited $n$-state, on the other hand, an extra pre-injection/preparatory scheme/mechanism is needed, which could potentially be a dye laser with a high repetition (MHz) rate, but one should stress that in a general context several schemes of this sort were discussed in the literature recently (for schemes based on adiabatic passage see~\cite{Vitanov,Vewinger,Vitanov2,Bergmann-review} and for pump-probe schemes see e.g.~\cite{Yuan}). After the injection, the survival of the atoms in their excited states is an issue that should to be considered. However, typically the lifetime of Rydberg atoms due the spontaneous decay or background microwave ionization is in the microsecond (10$^{-6}$ s) range \cite{gallagher2005rydberg}, i.e.~much longer than the strong field ultrashort processes considered here, typically developed in a sub-fs (10$^{-15}$ s) time scale.  Moreover, Bleda {\it et al.}~\cite{bleda2013high} have shown that the field-effect ionization rate dramatically decreases with increasing $n$. Thus, the strongest contribution to the ionization and HHG process comes directly from the initially prepared excited $n$-state atoms. Accordingly, we could anticipate that the Rydberg atoms, injected in the nano-gap region, can largely survive during the strong field interaction process.    

We employ the numerical solution of the TDSE, both in 1D and 3D, to compute the HHG spectra of an atom in a plasmon-enhanced linearly-polarized laser field. To avoid any misunderstanding we stress that this paper is not about determination of the plasmonic-enhanced fields, which is another important and challenging problem. Here we assume a certain spatial form and strength for the plasmonic-enhanced fields and they are treated as a given external field. Still these fields can and are controlled in experiments by changing the laser intensity, wavelength, polarization, or by fine-tunning the properties and geometry of the metallic nanostructure.

In this paper we systematically study the atomic response, when we alter the initial target atom bound state.  We increase the principal quantum number of the initial Rydberg state $n$,  and $n$-scale the relevant field parameters accordingly. In our simulations, we assume a bow-tie shaped nano-antenna as our plasmonic-enhanced field source, since this particular case has been extensively investigated~\cite{schuller2010plasmonics,kim2008high,park2011plasmonic,sivis2013extreme, pfullmann2013bow,park2013generation}. However, our simulations could be in principle extended to any nano-structure element whose spatio-temporal profile is analogous to the bow-tie nano-antennae one. At the same time we stress that in the explicit calculations we use a "caricature" of the true spacial dependence of the plasmonic-enhanced field, assuming that locally in the vicinity of the considered atom, it depends linearly on the space coordinates. It should be stressed, however, that this simple approximation grabs the main effects of the electric field inhomogeneity in space; for more careful description of plasmonic 
enhanced fields see for instance Refs.~\cite{ciappina2012optexp,ilhan2015h2p}, where finite differences time dependent (FDTD) codes were used to solve the Maxwell equations to determine the accurate shape and form of the spatially inhomogeneous fields.                                  
 
 We stress also that here the calculations are restricted to $n \leqslant 8$ since the span of the electron wave-packet (quiver radius) is on the order of the gap dimension ($\sim 20$ nn). For $n > 8$, on the other hand, the continuum electron could reach the metal surfaces and be absorbed. Finally, we assume that the field-enhancement factor is kept constant as the field intensity and laser frequency are $n$-scaled.    

The paper is organized as follows. In Sec.~II, we describe our theoretical methodologies for simulating the HHG process from Rydberg atoms driven by plasmonic-enhanced fields. In Sec.~III we present results using a 1D model atom. On the one hand, in Sec. III A, we simulate the electron wavepacket dynamics and obtain plasmonic-HHG from Rydberg atoms with $\it{fixed}$ laser field parameters (i.e.~unscaled field amplitude $E_0$ and unscaled frequency $\omega_0$). On the other hand, in Sec.~III B, the simulations are performed for $n$-scaled $E_0$ and $\omega_0$. Finally, in Sec.~III C, we compare the HHG yields obtained both in Secs.~III A and III B and discuss their main features, similarities and differences. Sec.~IV is focused on the TDSE simulations in 3D. Here, we model an H atom interacting with a plasmon-enhanced laser field and compute the HHG spectra for different laser parameters and excited states. In order to complete the analysis, in Sec.~V we perform semi-classical simulations, based on the three-step model. We conclude our contribution in Sec.~VI. 

\section{Methodology}

We investigate the mechanism of plasmonic-HHG from Rydberg atoms through the numerical solution of the time-dependent Schr\"odinger equation (TDSE). The interaction of a target atom with a plasmon-enhanced linearly polarized laser field is modelled via two different approaches; a 1D model atom and a real Hydrogen atom in 3D. Below we provide the details of these two methodologies.

\subsection{Model Hydrogen in 1D}

The TDSE describing the interaction of a 1D model atom with a laser field is written as follows (atomic units will be used throughout the article unless otherwise stated):
\begin{equation}
\label{tdse1d}
i\frac{\partial }{\partial t} \psi (x,t) = \left [ -\frac{1}{2} \frac{\partial^2 }{\partial x^2} + V(x)+x E(x,t) \right] \psi (x,t).
\end{equation}

Here the model soft-core potential is taken as $V(x)=-1/\sqrt{2+x^2}$. From the field-free solutions $\psi_n(x)$ of the Schr\"{o}dinger equation for the 1D model atom, one can find that the ionization potential of the ground state is $I_p=0.5$ a.u. and the energy levels follow 
\begin{equation}
E_n = \frac{-a}{(n+b)^2}, 
\end{equation} where the parameters $a$ and $b$ can be found by fitting the time-independent solutions of Eq.~(\ref{tdse1d}). Although, only the ground-state energy matches that of a real Hydrogen atom, a Rydberg-like character of the bound-state energies for high $n$ values (i.e.~$E_n \sim n^{-2}$ for $n \gg 1$) implies that the scaling rules discussed in Sec.~I are still valid for this 1D model atom~\cite{su1991model}. Taken the center of the nano-volume as the coordinate origin, the spatio-temporal profile of the laser field, represented by $E(x,t)$, is assumed to be in the following form~\cite{husakou2011theory,yavuz2012generation,ciappina2012high}

\begin{equation}
\label{efield1d}
E(x,t)\simeq E(t)[1+ h(x)],
\end{equation} 
where the space-free portion of the electric field is $E(t)= E_0 f(t) \cos(\omega_0 t)$. $E_0$ and $\omega_0$ are the peak amplitude ($E_0 = \sqrt{I/I_0}$ (a.u.) with $I_0 = 35.1$ PW/cm$^2$) and the frequency of the driving laser electric field, respectively. $f(t)$ defines the pulse envelope and is taken as a flat-top shape with 20 cycles long with one cycle ramp up and down. In Eq.~(\ref{efield1d}) $h(x)$ represents the functional form of the plasmonic-enhanced field. On the main advantages of the 1D model is that it allows to include any functional form for $h(x)$ (for different examples see e.g.~\cite{ciappina2012high,ciappina2012optexp,shaaran2013high}). Nevertheless, it is often sufficient to approximate $h(x)$ by a linear dependence, i.e.~$h(x) \simeq \beta x$, where $\beta$ defines the region of spatial inhomogeneity of the field. Therefore, $\beta$ has the units of inverse length. Results have shown that $\beta$ is an instrumental parameter to control the laser induced dynamics of the ionized electrons and, consequently, the modifications observed in the spectral profile of plasmonic-HHG~\cite{husakou2011theory,yavuz2012generation,ciappina2012high}.

\subsection{Real Hydrogen in 3D}

We also employ the numerical solution of the TDSE of an Hydrogen atom in 3D interacting with a linearly-polarized, in the $z$-axis, plasmon-enhanced laser field. The TDSE in the length gauge can be written as: 
\begin{equation}
i\frac{\partial \psi(\mathbf{r},t)}{\partial t}=\left[-\frac{\nabla^2}{2} +V(r)+ z E(z,t)\right]\psi(\mathbf{r},t),
\end{equation}
where $V(r)=-1/r$ is the atomic potential for the H atom. Here, the spatio-temporal profile of the laser field is taken to be similar to that we used for the 1D model atom, i.e.~$E(z,t)=E_0 f(t) (1+ \beta z)\cos(\omega_0t)$. However, this time we use a $\sin^2$-shaped pulse with four-cycles of total duration. The details of the 3D-TDSE numerical solution for an H atom in a plasmon-enhanced laser field can be found in Refs.~\cite{yavuz2012generation,fetic2013high}. In addition, the 3D-TDSE is able to model with precision any atom within the single active electron (SAE) approximation (see e.g.~\cite{perez2013beyond} for the He atom) by tunning adequately the atomic potential $V(r)$. Two disadvantages of the 3D-TDSE model are, (i) the high computational cost and (ii) the complications to model a general spatial shape for the plasmonic-enhanced field (to the best of our knowledge only the linear case has been modeled). Since the energy levels of the H atom scales as $n^{-2}$, the scaling laws provided in Sec.~I exactly applies. 

For both the 1D and 3D simulations, boundary reflection mask functions of the form $\cos^{1/8}$ multiply the electron wavefunction at each time-step in order to avoid spurious reflections~\cite{krause1992calculation}. The HHG spectra are then calculated from the modulus-square of the Fourier transform $a(\omega)$ of the dipole acceleration $a(t)$~\cite{burnett1992calculation}. Furthermore, the spatial dependence of the field-enhancement in Eq.~(\ref{efield1d}) is known to be an approximate expression against the realistic field-enhancement~\cite{husakou2011theory,ciappina2012optexp,shaaran2013high}. However, in many theoretical studies, Eq.~(\ref{efield1d}) has been shown to be sufficient to elucidate the underlying physics of the problem of interest~\cite{husakou2011theory,yavuz2012generation,ciappina2012high,perez2013beyond,fetic2013high,ciappina2013high,park2013generation,yavuz2013gas,zhang2013control,he2013wavelength,luo2014efficient,ebadi2014interferences}.         

\section{Plasmonic-HHG from Rydberg states of a 1D model atom}

\subsection{Plasmonic-HHG with unscaled field parameters}

\begin{figure}[ht]
\begin{flushleft} 
\includegraphics[width=0.54\textwidth,left]{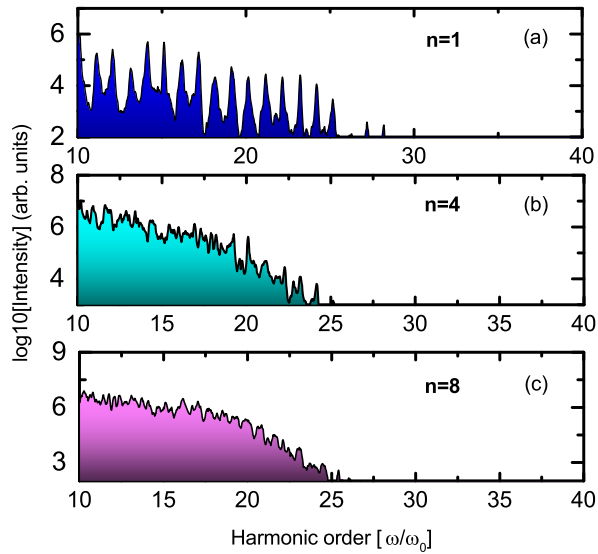}
\caption{\label{fig1} (color online) Plasmonic-HHG from $n$-states obtained from the 1D model atom. Here, the $n=1, 4$ and $8$ states are presented. $I = 20$ TW/cm$^2$ (1 TW/cm$^2$ = $1\times10^{12}$ W/cm$^2$) and $\lambda=800$ nm are used in all cases. The parameter $\beta$ is chosen as 0.016 a.u.}
\end{flushleft} 
\end{figure}

In this section, we examine plasmonic-HHG from the Rydberg series of a 1D model atom. We employ $n=1-8$ for an $\it{unscaled}$ intensity $I$ and frequency $\omega_0$. Here we take $I=20$ TW/cm$^2$ (1 TW/cm$^2$ = $1\times10^{12}$ W/cm$^2$) and $\lambda = 800$ nm, i.e.~$\omega_0 = 0.057$ a.u. (photon energy 1.55 eV). The field-inhomogeneity parameter $\beta$ is taken as 0.016 a.u., corresponding to a factor of $\sim 15$ intensity enhancement near the metallic surface, i.e.~$10$ nm away from the gap center, which is similar to that reported by Kim et al.~\cite{kim2008high} (note that this is an additional enhancement on top of the increase generated by the surface plasmon polaritons (SPP)). 

For demonstration purposes, Figs.~\ref{fig1}(a)-\ref{fig1}(c) show results only for $n=1, 4$ and $8$, respectively. First of all, for $\beta = 0.0$, the cut-off position for $n=1$ state is found to be $q_{\mathrm{cut}}=11$ (which corresponds to an $\omega_{\mathrm{cut-off}}=17.4$ eV) using the three-step model~\cite{corkum1993plasma}. As shown in Fig.~\ref{fig1}(a), the HHG cut-off is extended by roughly a factor of 2 to $q_{\mathrm{cut}}=23$ (corresponding to an $\omega_{\mathrm{cut-off}}=35.6$ eV), when the field inhomogeneity parameter is $\beta=0.016$. The increase in the HHG cutoff with $\beta \neq 0.0$ is consistent with previous theoretical studies~\cite{husakou2011theory,yavuz2012generation,ciappina2012high,perez2013beyond,fetic2013high,ciappina2013high,park2013generation,yavuz2013gas,zhang2013control,he2013wavelength,luo2014efficient,ebadi2014interferences}. 

\begin{figure}[hb]
\vspace{-10.0mm}
\includegraphics[width=0.52\textwidth,left]{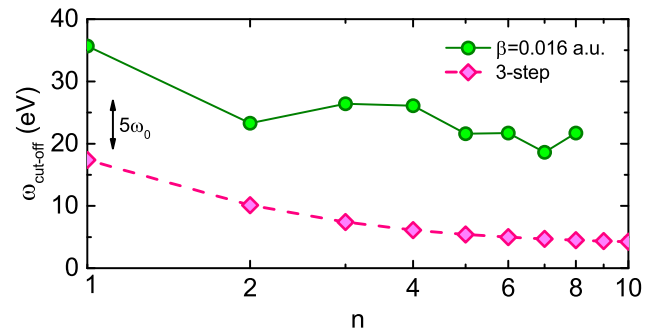}
\vspace{-17.0mm}
\caption{\label{fig2} (color online) Variation of $\omega_{\mathrm{cut-off}}$ of plasmonic-HHG ($\beta=0.016$ a.u.) with $n$ for fixed values of $E_0$ and $\omega_0$. The dashed line shows the predictions of the three-step model.}  
\end{figure}

The increase in the HHG cutoff originates from the further acceleration of the ionized electron moving in the plasmon-enhanced laser field. This results in a ponderomotive potential boost, due to the increase in the field strength as a function of increasing the spatial coordinate $x$~\cite{husakou2011theory,yavuz2012generation}. Since the highest harmonic photon energy depends on $I_p$ and $U_p$~\cite{corkum1993plasma}, the larger $U_p$, a higher $\omega_{\mathrm{cut-off}}$ will be reached. One could also see the appearance of both odd and even harmonics up to the HHG cutoff and below, which is attributed to the breaking of inversion symmetry owing to the spatial inhomogeneity of the plasmonic-enhanced field~\cite{husakou2011theory,ciappina2012high}. Beyond $n=1$ we find that the profile of the harmonic spectrum is very similar for $n=2-8$, namely they appear to be independent of the $n$-state (the $n=4$ and $8$ cases are shown in Figs.~\ref{fig1}(b) and \ref{fig1}(c), respectively). 

In Fig.~\ref{fig2} we depict the maximum photon energy ($\omega_{\mathrm{cut-off}}$) by varying the $n$-state for both inhomogeneous ($\beta=0.016$ a.u.) and homogeneous fields ($\beta=0.0$). $\omega_{\mathrm{cut-off}}$ drops significantly with $n$ for $\beta=0.0$ and converges at an $\omega_{\mathrm{cut-off}}=3.8$ eV, which corresponds to $3.17U_p$ ($U_p=1.19$ eV). In contrast, beyond the $n=1$ state, $\omega_{\mathrm{cut-off}}$ oscillates with a narrow energy band around $5 \omega_0$ above the conventional case for $\beta=0.016$ a.u. The modest dependence of $\omega_{\mathrm{cut-off}}$, as well as the conversion efficiency, to the $n$-state ($n>1$) for $\beta \neq 0$, could be attributed to the ionization dynamics and electron confinement during propagation, which strongly correlate with the maximum photon energy and harmonic efficiency~\cite{corkum1993plasma}. 

\begin{figure}[ht]
  \centering
  \includegraphics[width=0.48\textwidth,left]{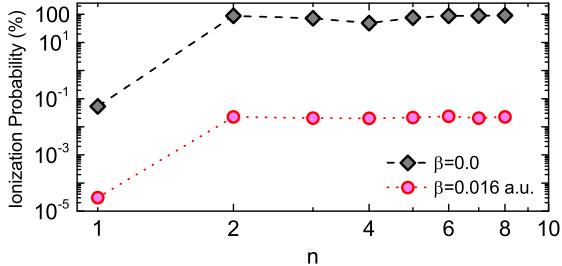}
\vspace{-25.0mm}
\caption{\label{fig3} (color online) Ionization probability as a function of the $n$ state for $\it{unscaled}$ field parameters, $E_0$ and $\omega_0$. Note that the ionization saturates beyond $n=2$ for both cases. Due to the possibility of electron confinement via repulsive forces (towards the nucleus) the ionization is largely suppressed in all $n$ cases with $\beta=0.016$ a.u.}
\end{figure}

For a classical electron of an hydrogen-like atom in a circular orbit,  
\begin{equation}
\label{Ibsi}
I_{BSI}=6.02 \times 10^{3} {\left | E_n [\mathrm{a.u.}] \right |}^3\, \mathrm{TW/cm}^2
\end{equation} determines the threshold intensity for the barrier-suppression-ionization (BSI), a critical point in which the Coulomb barrier is suppressed below the bound-state $n$ with energy $E_n$~\cite{bauer1997ejection,shakeshaft1990multiphoton}. 

The intensity value $I=20$ TW/cm$^2$ we use for $n=1$ and $\beta=0.0$ is much below $I_{BSI}$ ($I_{BSI}=750$ TW/cm$^2$), but it is above for other $n$ states since the $I_{BSI}$ condition in Eq.~(\ref{Ibsi}) scales as $n^{-6}$. Fig.~\ref{fig3} supports this argument. We observe that the ionization saturates beyond $n=1$ (with almost full ionization) for $\beta=0.0$. Thus, one should not expect an efficient HHG from $n \geq 2$ states with homogeneous fields, because of the likelihood of no electron rescattering. 

However, note that the condition given in Eq.~(\ref{Ibsi}) is valid only for homogeneous fields, i.e.~for the case of $\beta=0.0$. Through numerical simulations we find that when the atom is placed in a plasmonic-enhanced field, the ionization probabilities are suppressed by roughly four orders of magnitude (see Fig.~\ref{fig3}), owing to the electron confinement via repulsive forces (towards the nucleus). Thus, the confined continuum electrons -which are likely to escape and never return, in the case of homogeneous fields- could be driven back, resulting in an efficient recombination with the parent ion, leading to an ultimate coherent photo-emission. 

Although our results appear to be promising, namely, one could attain more energetic and efficient photons through a Rydberg-series atom driven by a plasmonic-enhanced laser-field, the laser field intensity increases in such a way to exceed the damage threshold of the metallic nanostructure and thus this procedure is not very beneficial (see Sec.~I). In order to give a clearer picture, in the next section, we investigate plasmonic-HHG from a Rydberg-series atom for a set of $scaled$ laser field parameters.     

\subsection{Plasmonic-HHG with n-scaled field parameters}

Here we study the plasmonic-HHG from Rydberg atoms by systematically $n$-scaling the input parameters. As we go up in $n$, the field intensity $I$ and frequency $\omega_0$, are scaled as $20n^{-8}$ TW/cm$^2$ and $1.55n^{-3}$ eV, respectively. We assume that the spatial dependence of the plasmonic-enhanced field is fixed, independent of the value of $n$. Although, the field frequency is critically affecting the plasmonic-field enhancement, we assume that as we $n$-scale the field frequency, the degree of field enhancement through surface plasmon resonances are also fixed. Nonetheless, this condition could potentially be fulfilled by tweaking the configuration and geometry of the nanostructure element systematically as to maintain the degree of the field enhancement at different field wavelengths~\cite{barnes2003surface,eustis2006gold,willets2007localized}.

\begin{figure}[hb]
\centering
\includegraphics[width=0.54\textwidth,left]{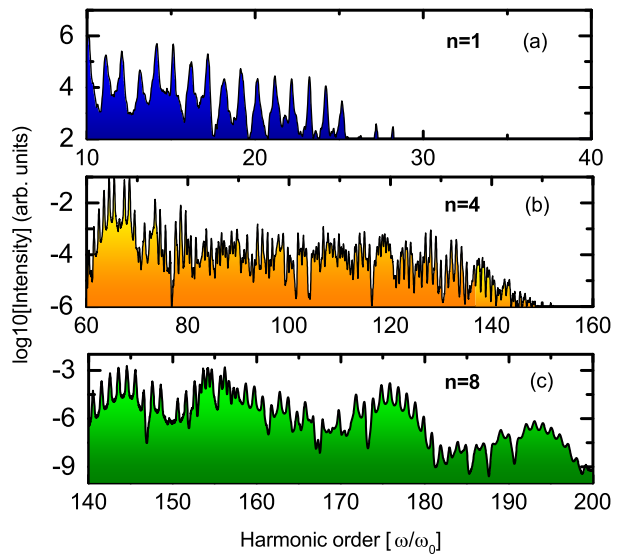}
\caption{\label{fig4} (color online) Plasmonic-HHG from $n$-states of the 1D model atom. The field parameters are scaled as $I = 20/n^8$ TW/cm$^2$ and $\lambda=800n^3$ nm for each $n$ state. The field inhomogeneity $\beta$ is chosen as 0.016 a.u.}
\end{figure}

Thus, for higher values of $n$, the intensity enhancement in the nanostructure volume would be much lower and, as a result, reaching the nanostructure damage threshold is highly unlikely. It has been shown that $100-1000$ TW/cm$^2$ of peak intensities are harmful to nanostructures~\cite{kim2008high,park2011plasmonic,sivis2013extreme, pfullmann2013bow}, therefore alternatives to diminish these values would be highly desirable. These values are also valid for our set of $I$ and $\beta$ parameters, since the intensity is enhanced up to $300$ TW/cm$^2$ near the metal's surfaces. However, we could use, for instance, $I=0.3$ GW/cm$^2$ for the $n=4$ state. In addition, the intensity $I$ near the metal surfaces results $4.5$ GW/cm$^2$, which is still considerably below the damage threshold. Note that this maximum value depends on the spatio-temporal shape of the plasmonic-enhanced electric field function and the field inhomogeneity parameter $\beta$ we use. In Fig.~\ref{fig4}, we present results for $n=1, 4$ and $8$. As can be observed more harmonics $\omega/\omega_0$ are covered with increasing $n$. 

There are clear cutoffs at the 23$^{\mathrm{th}}$, 138$^{\mathrm{th}}$ and 193$^{\mathrm{th}}$ harmonics for $n=1, 4$ and $8$ states, respectively. Note also that, a more regular and clear harmonic peaks for $n=4$ and $8$ are obtained, as opposed to the irregular ones observed in Fig.~\ref{fig1}. Although the order of the cutoff position is increased, maximum photon energies are decreased at the value of 35.6 eV, 3.3 eV and 0.6 eV, for $n=1$, $4$ and $8$ states, respectively. This is so, since the photon energies are given as $\omega = q \omega_0$ and $\omega_0$ scales as $n^{-3}$ as in Eq.~(\ref{omegascal}). 

\begin{figure}[ht]
\centering
\vspace{-10.0mm}
\includegraphics[width=0.51\textwidth,left]{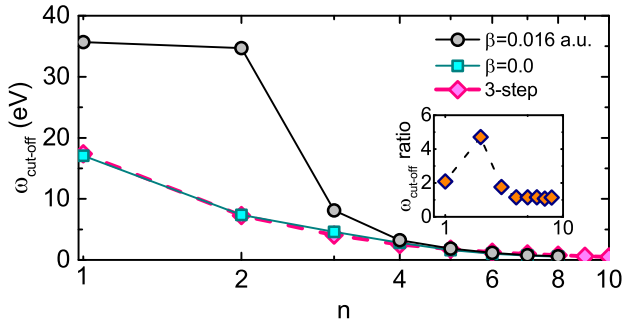}
\vspace{-15.0mm}
\caption{\label{fig5} (color online) The variation of $\omega_{\mathrm{cut-off}}$ of plasmonic-HHG ($\beta=0.016$ $a.u.$) and HHG ($\beta=0.0$) with $n$ for $n$-scaled $E_0$ and $\omega_0$. The dashed line shows predictions of the three-step model. Inset shows $\omega_{\mathrm{cut-off}}$ enhancement with respect to $n$. }
\end{figure}

The variation of the $\omega_{\mathrm{cut-off}}$ with $n$ (for $n=1-8$) for both homogeneous and inhomogeneous laser fields is shown in Fig.~\ref{fig5}. As can be inferred, the $\omega_{\mathrm{cut-off}}$ for homogeneous laser fields decreases with $n$. This decline scales as $n^{-2}$ since $I_p \sim n^{-2}$ for higher values of $n$ in our 1D model atom. Our results of $\omega_{\mathrm{cut-off}}$ for homogeneous laser fields match exactly the predictions of the three-step model, i.e.~$\omega_{\mathrm{cut-off}}=I_p + 3.17 U_p$~\cite{corkum1993plasma,lewenstein1994theory}.

We also shown in Fig.~\ref{fig5} the variation of $\omega_{\mathrm{cut-off}}$ with $n$ for an atom exposed to a plasmon-enhanced field with $\beta=0.016$ a.u. We see from Eq.~(\ref{efield1d}) that the spatial intensity enhancement becomes effective once the electron wavepacket is released and pushed away from the nucleus. In other words, $(1+\beta x)\rightarrow 1$ for $x \rightarrow 0$. Therefore, to a first-order approximation, we can employ the conventional quasi-static ionization rate expressions to understand the influence of $n$ on ionization. Quasi-static ADK ionization rates are defined by~\cite{ammosov1986tunnel}
\begin{equation}
\Gamma \sim \exp[-2(2I_p)^{3/2}/(3E_0)],
\end{equation} 
therefore, it is clear that the [$(2I_p)^{3/2}/(3E_0)$] term scales as $n$. As a result the tunnel ionization probability drops drastically with increasing $n$. We can then conclude that the enhancement in $\omega_{\mathrm{cut-off}}$ with $n$, caused by the field inhomogeneity, would decrease for $n \gg 1$. Fig.~\ref{fig5} shows that there is a dramatic increase in $\omega_{\mathrm{cut-off}}$ for $n=1$ and $n=2$ and a small increase for $n=3$, over those of the homogeneous laser field case. However, in agreement with our previous statements, the influence of the field inhomogeneity considerably drops for $n > 3$. 

\begin{figure}[ht]
\centering
\vspace{-10.0mm}
\includegraphics[width=0.50\textwidth,left]{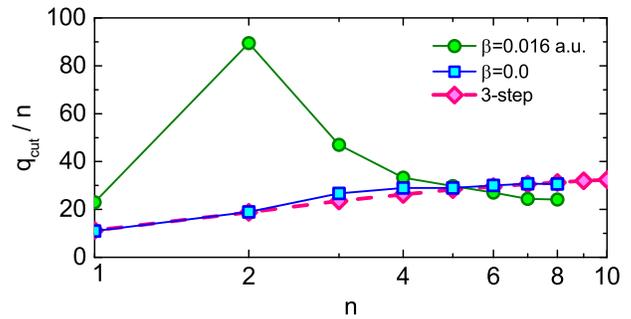}
\vspace{-15.0mm}
\caption{\label{fig6} (color online) Variation of the scaled cutoff position $q_{\mathrm{cut}}/n$ with $n$. Dashed-line and $\Box$ represent the three-step model's predictions and our TDSE results for the homogeneous field, $\beta=0.0$, respectively. $\bigcirc$ represents the $q_{\mathrm{cut}}/n$ ratio for plasmon-enhanced inhomogeneous fields for $\beta=0.016$ a.u. The points are connected with lines for tracking with ease.}
\end{figure}

For practical reasons, we examine the $n$-scaling of the HHG cutoff position $q_{\mathrm{cut}}$ with respect to the bound-states $n=1-8$ of our 1D model atom. Using the three-step model for the Rydberg-like series, with the $n^{-2}$ bound-state scaling, we observe that $\omega_{\mathrm{cut-off}}$ scales as $n^{-2}$, therefore the HHG cutoff position results $q_{\mathrm{cut}}=\omega_{\mathrm{cut-off}}/\omega_0 \sim n$~\cite{bleda2013high}. Then we can deduce that the scaled HHG cutoff position can be written as
\begin{equation}
\label{qcut}
\frac{q_{\mathrm{cut}}}{n}=\mathrm{const},
\end{equation} 
which should exactly be satisfied for $E_n \sim n^{-2}$~\cite{bleda2013high}. Fig.~\ref{fig6} shows the variation of the scaled cut-off ($q_{\mathrm{cut}}/n$ ratio) for homogeneous ($\beta=0.0$) and inhomogeneous fields (with $\beta=0.016$ a.u.). As can be seen, the scaled HHG cut-off for $\beta=0.0$ increases with $n$ and converges to a constant value in agreement with Eq.~(\ref{qcut}). In our simulations we find that, in agreement with the three-step model, $q_{\mathrm{cut}}/n \rightarrow 32$ for $n \gg 1$. For $\beta=0.016$ a.u., however, the condition provided in Eq.~(\ref{qcut}) is not fulfilled, due to the non-ponderomotive character of the electron acceleration~\cite{herink2012field}. On the other hand, although for $\beta=0.016$ a.u.~we stated that the influence of the nanostructure element diminishes for higher values of $n$, we can clearly see here that beyond $n=5$ the cut-off position is close but begin to deviate slightly from those predicted by the three-step model and/or our TDSE simulations for $\beta=0.0$. This behavior arises because the initial width of the wavepacket approaches the metal's surfaces for higher values of $n$, increasing the chances of surface-absorption of energetic electrons with high linear momentum~\cite{husakou2011theory,yavuz2012generation,yavuz2013gas}.  

\subsection{$n$-scaling of plasmonic-HHG yield}

So far we have demonstrated HHG from a Rydberg atom in inhomogeneous laser fields for constant or $n$-scaled field parameters, i.e.~$E_0$ and $\omega_0$. However, one can ask the question, which of these two procedures is more beneficial in terms of conversion efficiency? This is investigated in terms of the interplay amongst the maximum photon energies, their yields and probability of reaching the nanostructure material damage threshold.   

\begin{figure}[ht]
\centering
\includegraphics[width=0.48\textwidth,left]{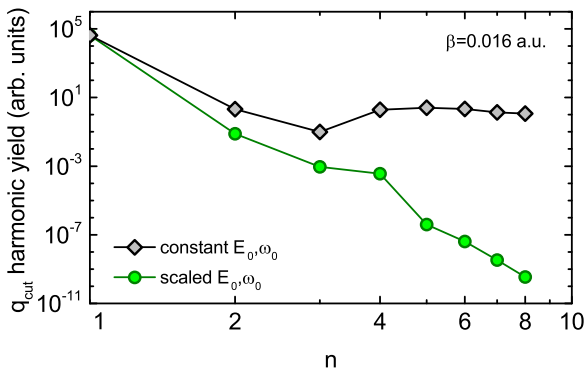}
\vspace{-12.0mm}
\caption{\label{fig7} $n$-scaling of the harmonic efficiency of $q_{\mathrm{cut}}$ of $\beta=0.016$ a.u.~for the values presented in Fig.~\ref{fig2} and Fig.~\ref{fig5} for constant and scaled field parameters, respectively.} 
\end{figure}

It has been noted that, for conventional (homogeneous) strong-field interactions, the efficiency of harmonics in HHG is inversely proportional to the degree of the electron wavepacket's transverse spread at the time of recombination with the parent ion~\cite{delone1991energy,lewenstein1994theory,ivanov1996coulomb},

\begin{equation}
\label{sigmax1}
\sigma_x\propto \frac{E_{0}^{1/2}}{I_p^{1/4}{\omega_0}}.
\end{equation}As we go up in $n$, the wavepacket spread scales as    

\begin{equation}
\label{sigmax2}
\sigma_{x} = n^{3/2} \widetilde{\sigma}_{x},
\end{equation} 
for fixed $E_0$ and $\omega_{0}$, and 
\begin{equation}
\label{sigmax3}
\sigma_{x} = n^{1/2} \widetilde{\sigma}_{x},
\end{equation}
for $n$-scaled $E_0$ and $\omega_{0}$. We can see that the electron wavepacket spread increases, thus the efficiency of harmonics for homogeneous laser fields drops with $n$ for both scaled and constant field parameters. However, the decay is relatively faster for the scaled ones. In Fig.~\ref{fig7}, the same asseveration applies for plasmon-enhanced laser fields, although the variation in harmonic efficiency of $q_{cut}$ is not very regular. For constant laser field parameters, the harmonic yield drops and saturates after $n=3$, in accordance with what is observed in Fig.~3. On the other hand, for scaled parameters, the decrease in yield is even more dramatic with $n$. 

There is clearly no benefit in using very high $n$ states, for plasmonic-HHG near the metallic nano-structure, in terms of maximum photon energy and their yields (see also Fig.~\ref{fig2} and \ref{fig5}). For the $n=2-8$ series with constant $E_0$ and $\omega_0$, the harmonic yields are relatively high, comparing with those of the scaled $E_0$ and $\omega_0$ (see Fig.~\ref{fig7}). As we previously stated, however, plasmon-enhanced field employed for these $n$ states are self-detrimental. We note that plasmonic-HHG from $n=2$ with scaled field parameters appears to be the best option amongst the whole studied series. 

\section{Plasmonic-HHG from a real H atom in 3D}

In this section, we would like to address whether some of the assessments on the plasmonic-HHG in a 1D model atom render in more realistic situations. For that, we perform simulations for the plasmonic-HHG from a real H atom in three-dimensions (3D).  

\begin{figure}[ht]
\centering
\vspace{-25.0mm}
\includegraphics[width=0.51\textwidth,left]{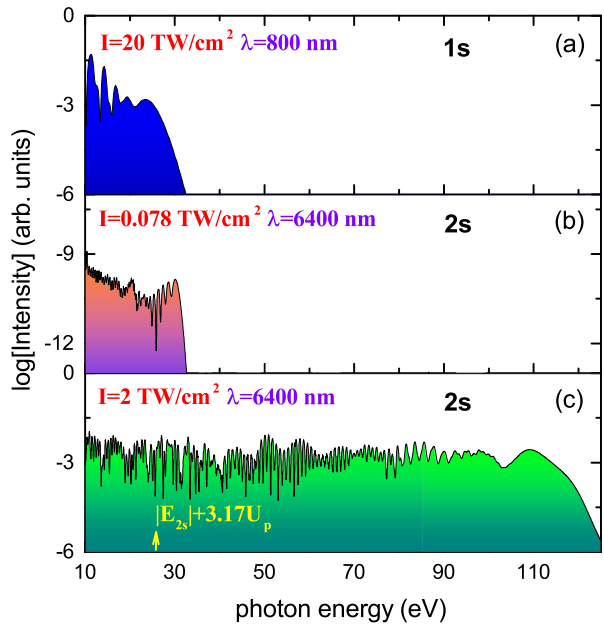}
\vspace{-15.0mm}
\caption{\label{fig8} (color online) Plasmonic-HHG from a real H atom in 3D. (a) Ground-state ($1s$) H exposed to a plasmon-enhanced field with a field inhomogeneity parameter $\beta=0.016$ a.u., intensity $I=20$ TW/cm$^2$ and $\lambda=800$ nm. (b) plasmonic-HHG from a H atom in the $2s$ state, again with $\beta=0.016$ a.u.. For the $2s$ state, the plasmonic field parameters are scaled to those of the panel (a), i.e.~$I$ is multplied by $2^{-8}$ and $\lambda$ by $2^{3}$. (c) plasmonic-HHG from a $2s$ state with $I=2$ TW/cm$^2$ and $\lambda=6400$ nm. The three-step model with $\beta=0.0$ predicts an HHG cutoff at $17$, $4.3$ and $28$ eV for (a), (b) and (c), respectively.} 
\end{figure}

In this Section we restrict our calculations to the ground ($1s$) and first excited states ($2s$), since, as we have shown in the Sec.~III, there might be no benefit in going beyond the $n=2$ state in terms of plasmonic-HHG efficiency and maximum photon energy. Figs.~\ref{fig8}(a)-\ref{fig8}(c) show the results of our simulations. We first use $I=20$ TW/cm$^2$ and $\lambda=800$ nm for an H atom in a $1s$ state (Fig.~8(a)) and scale these parameters with $n^{-8}$ and $n^{3}$ for the $2s$ state case, respectively. Thus, the field parameters become $I=0.078$ TW/cm$^2$ and $\lambda=6400$ nm, respectively (Fig.~\ref{fig8}(b)). For these two cases we use $\beta=0.016$ a.u.~and the laser intensities are enhanced up to $300$ TW/cm$^2$ and $1.2$ TW/cm$^2$ near the metal surfaces, respectively. For the selected laser field parameters, the maximum photon energies predicted by the three-step model, for the $1s$ and $2s$ states, are $17$ eV and $4.35$ eV, respectively. However, as shown in Fig.~\ref{fig8}(a), when the H atom in the ground state is exposed to a plasmonic-enhanced field with $\beta=0.016$ a.u., the plateau is extended up to $\sim 23$ eV. This value is slightly smaller than that of the 1D model atom in the ground state (see Fig.~\ref{fig5}). This could be attributed to the limitations in the 1D model to describe the transversal spreading of the electron wavepacket. Note that the binding-energy (so is the $I_p$) of the 1D model atom in the ground state is the same as the real H atom. For the $2s$ state, shown in Fig.~\ref{fig8}(b), we observe a plasmonic-HHG extended from a theoretically predicted value of $4.35$ eV (for $\beta=0.0$) to $30$ eV. However, comparing with the $1s$ state case, the efficiency of the plateau drops almost seven orders of magnitude. This drop is a consequence of an increase in the transversal spreading of the electron wavepacket in the continuum when going up from the $1s$ to the $2s$ state (see Eqs.~(\ref{sigmax1})-(\ref{sigmax3})). 

Finally, we demonstrate a plasmonic-HHG from a $2s$ state using a laser field whose intensity is stronger than $I=0.078$ TW/cm$^2$, but still well below the damage threshold. Here we use $I'=2$ TW/cm$^2$ for an H atom in a $2s$ state with $\lambda=800\cdot2^3=6400$ nm (6.4 $\mu$m) of wavelength (the corresponding $\omega_0=0.194$ eV) (Fig.~\ref{fig8}(c)). The three-step model predicts a value of $\omega_{\mathrm{cut-off}}=26$ eV for an homogeneous laser field. As shown in Fig.~\ref{fig8}(c), a 25-fold increase in intensity results in roughly 4-fold extension in the plateau region for this $2s$ state. Moreover, the efficiency of the plasmonic-HHG approaches to that of the $1s$ state (see Fig.~\ref{fig8}(a)). This example clearly shows the benefit of employing Rydberg atoms in plasmonic-HHG. However, in general, the interplay between the propagation time and spatial extension of the electron wavepacket, joint with the possibility of absorption of the continuum electron at metal surfaces during its excursion should to be considered. Furthermore, the means of avoiding the damage to the nanostructure element are amongst the most crucial phenomena to exploit the maximum efficiency of plasmonic-HHG from Rydberg atoms.

\section{Semi-classical Plasmonic-HHG}

In this section, we perform semi-classical simulations extended to atoms in a plasmon-enhanced field~\cite{husakou2011theory,yavuz2012generation,ciappina2012high,ciappinacpc}. In order to simulate Rydberg atoms we consider two cases; the electron is released either from a $n=1$ state or a $n=2$ state. For $n=2$, the laser parameters and the energy levels are scaled to those of $n=1$. The motion of the semi-classical electron under the influence of a plasmonic laser field is described by
\begin{equation}
\ddot{x}(t)=(1+2\beta x)E(t),
\end{equation} 
where $E(t)$ is defined in Sec.~II. The semi-classical equations are numerically solved by velocity Verlet algorithm. The electron is released at time $t_i$ and position $x_n(t_i)=3n^2/2$ with a zero kinetic energy and travels under the influence of the electric field. Here, $x_n$ is the expectation value of the semi-classical electron bound to the $n^{\mathrm{th}}$ state of an H atom. On the other hand, due to the possibility of electron absorptions at metal surfaces, we make sure that the trajectories that reach the metallic surfaces are neglected~\cite{husakou2011theory}. For those electrons which return back and recombines with the parent ion at time $t_r$, the total energy is then $|E_n|+E_k(t_r)$, where $E_n=-0.5/n^2$ is the energy of the $n^{\mathrm{th}}$ level of the H atom and $E_k(t_r)$ is the kinetic energy of the recombined electron. 

\begin{figure}[ht]
\centering
\vspace{0.0mm}
\includegraphics[width=0.51\textwidth,left]{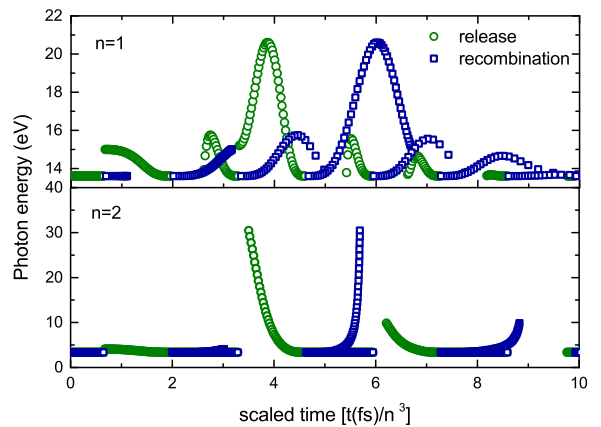}
\vspace{-10.0mm}
\caption{\label{fig9} (color online) Semi-classical simulations for an electron in a plasmonic field. Graph shows the electron kinetic energy at the recombination as a function on the release and recombination times. It is assumed that the electron is released from $n=1$ (top figure) and $n=2$ (bottom figure) states. For $n=1$, $\beta=0.016$ a.u., intensity $I=20$ TW/cm$^2$ and $\lambda=800$ nm. For $n=2$, again $\beta=0.016$ a.u. and the plasmonic field parameters are correspondingly scaled to those of the top figure, i.e.~$I$ by $2^{-8}$ and $\lambda$ by $2^{3}$. The three-step model with $\beta=0.0$ predicts cutoffs at $17$, $4.3$ eV for $n=1$ and $2$, respectively.} 
\end{figure}

Figure \ref{fig9} shows the electron kinetic energy at the recombination, i.e.~the harmonic photon energy, in terms of harmonic order, as a function of the release (ionization) time $t_i$ and recombination $t_r$, for both $n=1$ and $n=2$. For $n=1$, the maximum energy is $21$ eV, which is in very good agreement with that predicted by the 3D TDSE (see Fig.~\ref{fig8}(a)) and lower than the prediction of the 1D model atom. For $n=2$, on the other hand, semi-classical simulations predict a maximum photon energy of $31.3$ eV. This value agrees well with the 3D TDSE prediction ($\sim 30$ eV) and, again, is slightly lower than the 1D TDSE case ($\sim 35$ eV). As was already mentioned, we could infer that the disagreement between the 1D and 3D TDSE models is due to the lateral spreading of the electron wavepacket, which is apparently missing in the 1D case. 

Although they are not shown here, we also find that the maximum photon energies decrease dramatically for $n>2$ states. This is consistent with the conclusions we have found for the 1D-TDSE case (see Fig.~\ref{fig5}). 

\section{Concluding remarks}

In conclusion, we have presented a complete theoretical investigation of the behavior of Rydberg atoms in plasmon-enhanced laser fields producing a, so-called, plasmonic-HHG. Since scaling is crucial in this context, as we increased the principal quantum number of the initial Rydberg state $n$, we have systematically $n$-scaled the other input parameters, such as the incident field strength and frequency, etc. We have simulated plasmonic-HHG based on the numerical solution of the TDSE for a 1D model atom and a real H atom in 3D. 

We have first simulated plasmonic-HHG from Rydberg atoms with fixed field intensity and laser frequency. We have found that in the case of homogeneous laser fields the dynamics of ionization is governed by the mechanism of barrier suppression (BSI) for $n>1$.  However, when we place the Rydberg atoms in an spatial inhomogeneous plasmon-enhanced field, we observe that ionization rates drop down several orders of magnitude. This is caused by the electron confinement into a nano-volume,  the recombination of the laser ionized electron is then possible, and therefore HHG plausible. This contrasts  with the case of Rydberg atoms in homogeneous fields where the recombination is strongly suppressed.

We have then $n$-scaled both the field intensity and frequency of the driving laser field as we go up in $n$. We have observed a maximum cutoff enhancement from the $n=2$ state. Beyond that point, the influence of the plasmonic-enhanced field is no longer significant. 

Finally, in order to complement our assessments, we have also employed the 3D TDSE for an H atom in a spatial inhomogeneous plasmon-enhanced field and simulated plasmonic-HHG from the ground ($1s$) and the first excited ($2s$) states. We have verified that the length of the plateau is on the same order for both $1s$ and $2s$ states, where the field parameters are $n$-scaled over those of the $1s$ state. Semi-classical simulations have revealed that the maximum photon energies for both the $n=1$ and $n=2$ states agree very well with the 3D simulations and lower than those of 1D, suggesting that lateral spreading of the wave-packet is critical in this process.             

\section{acknowledgment}

Y.T and Z.A acknowledge support from BAPKO of MU. Authors are grateful to Dr. E. A. Bleda for helpful discussions and a critical reading of the the manuscript.   M.L. and A. C. acknowledge the support of Spanish MINECO (Nacional Plan GRANT FOQUS FIS2013-46768-P, SEVERO OCHOA GRANT/SEV-2015-0522), Fundaci\'o Cellex,  Catalan AGAUR (SGR 2014 874) and ERC AdG OSYRIS.


\begin{thebibliography}{66}%
\makeatletter
\providecommand \@ifxundefined [1]{%
 \@ifx{#1\undefined}
}%
\providecommand \@ifnum [1]{%
 \ifnum #1\expandafter \@firstoftwo
 \else \expandafter \@secondoftwo
 \fi
}%
\providecommand \@ifx [1]{%
 \ifx #1\expandafter \@firstoftwo
 \else \expandafter \@secondoftwo
 \fi
}%
\providecommand \natexlab [1]{#1}%
\providecommand \enquote  [1]{``#1''}%
\providecommand \bibnamefont  [1]{#1}%
\providecommand \bibfnamefont [1]{#1}%
\providecommand \citenamefont [1]{#1}%
\providecommand \href@noop [0]{\@secondoftwo}%
\providecommand \href [0]{\begingroup \@sanitize@url \@href}%
\providecommand \@href[1]{\@@startlink{#1}\@@href}%
\providecommand \@@href[1]{\endgroup#1\@@endlink}%
\providecommand \@sanitize@url [0]{\catcode `\\12\catcode `\$12\catcode
  `\&12\catcode `\#12\catcode `\^12\catcode `\_12\catcode `\%12\relax}%
\providecommand \@@startlink[1]{}%
\providecommand \@@endlink[0]{}%
\providecommand \url  [0]{\begingroup\@sanitize@url \@url }%
\providecommand \@url [1]{\endgroup\@href {#1}{\urlprefix }}%
\providecommand \urlprefix  [0]{URL }%
\providecommand \Eprint [0]{\href }%
\providecommand \doibase [0]{http://dx.doi.org/}%
\providecommand \selectlanguage [0]{\@gobble}%
\providecommand \bibinfo  [0]{\@secondoftwo}%
\providecommand \bibfield  [0]{\@secondoftwo}%
\providecommand \translation [1]{[#1]}%
\providecommand \BibitemOpen [0]{}%
\providecommand \bibitemStop [0]{}%
\providecommand \bibitemNoStop [0]{.\EOS\space}%
\providecommand \EOS [0]{\spacefactor3000\relax}%
\providecommand \BibitemShut  [1]{\csname bibitem#1\endcsname}%
\let\auto@bib@innerbib\@empty
\bibitem [{\citenamefont {Corkum}\ and\ \citenamefont
  {Krausz}(2007)}]{corkum2007attosecond}%
  \BibitemOpen
  \bibfield  {author} {\bibinfo {author} {\bibfnamefont {P.~B.}\ \bibnamefont
  {Corkum}}\ and\ \bibinfo {author} {\bibfnamefont {F.}~\bibnamefont
  {Krausz}},\ }\href@noop {} {\bibfield  {journal} {\bibinfo  {journal} {Nature
  Phys.}\ }\textbf {\bibinfo {volume} {3}},\ \bibinfo {pages} {381} (\bibinfo
  {year} {2007})}\BibitemShut {NoStop}%
\bibitem [{\citenamefont {Brabec}\ and\ \citenamefont
  {Krausz}(2000)}]{brabec2000intense}%
  \BibitemOpen
  \bibfield  {author} {\bibinfo {author} {\bibfnamefont {T.}~\bibnamefont
  {Brabec}}\ and\ \bibinfo {author} {\bibfnamefont {F.}~\bibnamefont
  {Krausz}},\ }\href@noop {} {\bibfield  {journal} {\bibinfo  {journal} {Rev.
  Mod. Phys.}\ }\textbf {\bibinfo {volume} {72}},\ \bibinfo {pages} {545}
  (\bibinfo {year} {2000})}\BibitemShut {NoStop}%
\bibitem [{\citenamefont {Corkum}(1993)}]{corkum1993plasma}%
  \BibitemOpen
  \bibfield  {author} {\bibinfo {author} {\bibfnamefont {P.~B.}\ \bibnamefont
  {Corkum}},\ }\href@noop {} {\bibfield  {journal} {\bibinfo  {journal} {Phys.
  Rev. Lett.}\ }\textbf {\bibinfo {volume} {71}},\ \bibinfo {pages} {1994}
  (\bibinfo {year} {1993})}\BibitemShut {NoStop}%
\bibitem [{\citenamefont {Lewenstein}\ \emph {et~al.}(1994)\citenamefont
  {Lewenstein}, \citenamefont {Balcou}, \citenamefont {Ivanov}, \citenamefont
  {L'Huillier},\ and\ \citenamefont {Corkum}}]{lewenstein1994theory}%
  \BibitemOpen
  \bibfield  {author} {\bibinfo {author} {\bibfnamefont {M.}~\bibnamefont
  {Lewenstein}}, \bibinfo {author} {\bibfnamefont {P.}~\bibnamefont {Balcou}},
  \bibinfo {author} {\bibfnamefont {M.~Y.}\ \bibnamefont {Ivanov}}, \bibinfo
  {author} {\bibfnamefont {A.}~\bibnamefont {L'Huillier}}, \ and\ \bibinfo
  {author} {\bibfnamefont {P.~B.}\ \bibnamefont {Corkum}},\ }\href@noop {}
  {\bibfield  {journal} {\bibinfo  {journal} {Phys. Rev. A}\ }\textbf {\bibinfo
  {volume} {49}},\ \bibinfo {pages} {2117} (\bibinfo {year}
  {1994})}\BibitemShut {NoStop}%
\bibitem [{\citenamefont {Lewenstein}\ and\ \citenamefont
  {L'Huillier}(2009)}]{AnneML}%
  \BibitemOpen
  \bibfield  {author} {\bibinfo {author} {\bibfnamefont {M.}~\bibnamefont
  {Lewenstein}}\ and\ \bibinfo {author} {\bibfnamefont {A.}~\bibnamefont
  {L'Huillier}},\ }in\ \href@noop {} {\emph {\bibinfo {booktitle} {Strong Field
  Laser Physics}}},\ \bibinfo {editor} {edited by\ \bibinfo {editor}
  {\bibfnamefont {T.}~\bibnamefont {Brabec}}}\ (\bibinfo  {publisher}
  {Springer-Verlag},\ \bibinfo {address} {New York},\ \bibinfo {year} {2009})\
  pp.\ \bibinfo {pages} {147--183}\BibitemShut {NoStop}%
\bibitem [{\citenamefont {Gaarde}\ \emph {et~al.}(2008)\citenamefont {Gaarde},
  \citenamefont {Tate},\ and\ \citenamefont {Schafer}}]{Gaarde}%
  \BibitemOpen
  \bibfield  {author} {\bibinfo {author} {\bibfnamefont {M.~B.}\ \bibnamefont
  {Gaarde}}, \bibinfo {author} {\bibfnamefont {J.~L.}\ \bibnamefont {Tate}}, \
  and\ \bibinfo {author} {\bibfnamefont {K.~J.}\ \bibnamefont {Schafer}},\
  }\href@noop {} {\bibfield  {journal} {\bibinfo  {journal} {J. Phys. B}\
  }\textbf {\bibinfo {volume} {41}},\ \bibinfo {pages} {132001} (\bibinfo
  {year} {2008})}\BibitemShut {NoStop}%
\bibitem [{\citenamefont {Sanpera}\ \emph {et~al.}(1995)\citenamefont
  {Sanpera}, \citenamefont {J\"onsson}, \citenamefont {Watson},\ and\
  \citenamefont {Burnett}}]{Sanpera1995}%
  \BibitemOpen
  \bibfield  {author} {\bibinfo {author} {\bibfnamefont {A.}~\bibnamefont
  {Sanpera}}, \bibinfo {author} {\bibfnamefont {P.}~\bibnamefont {J\"onsson}},
  \bibinfo {author} {\bibfnamefont {J.~B.}\ \bibnamefont {Watson}}, \ and\
  \bibinfo {author} {\bibfnamefont {K.}~\bibnamefont {Burnett}},\ }\href@noop
  {} {\bibfield  {journal} {\bibinfo  {journal} {Phys. Rev. A}\ }\textbf
  {\bibinfo {volume} {51}},\ \bibinfo {pages} {3148} (\bibinfo {year}
  {1995})}\BibitemShut {NoStop}%
\bibitem [{\citenamefont {Tate}\ \emph {et~al.}(2007)\citenamefont {Tate},
  \citenamefont {Auguste}, \citenamefont {Muller}, \citenamefont {Sali\`eres},
  \citenamefont {Agostini},\ and\ \citenamefont {DiMauro}}]{Tate}%
  \BibitemOpen
  \bibfield  {author} {\bibinfo {author} {\bibfnamefont {J.}~\bibnamefont
  {Tate}}, \bibinfo {author} {\bibfnamefont {T.}~\bibnamefont {Auguste}},
  \bibinfo {author} {\bibfnamefont {H.~G.}\ \bibnamefont {Muller}}, \bibinfo
  {author} {\bibfnamefont {P.}~\bibnamefont {Sali\`eres}}, \bibinfo {author}
  {\bibfnamefont {P.}~\bibnamefont {Agostini}}, \ and\ \bibinfo {author}
  {\bibfnamefont {L.~F.}\ \bibnamefont {DiMauro}},\ }\href@noop {} {\bibfield
  {journal} {\bibinfo  {journal} {Phys. Rev. Lett.}\ }\textbf {\bibinfo
  {volume} {98}},\ \bibinfo {pages} {013901} (\bibinfo {year}
  {2007})}\BibitemShut {NoStop}%
\bibitem [{\citenamefont {Gohle}\ \emph {et~al.}(2005)\citenamefont {Gohle},
  \citenamefont {Udem}, \citenamefont {Herrmann}, \citenamefont
  {Rauschenberger}, \citenamefont {Holzwarth}, \citenamefont {Schuessler},
  \citenamefont {Krausz},\ and\ \citenamefont
  {H{\"a}nsch}}]{gohle2005frequency}%
  \BibitemOpen
  \bibfield  {author} {\bibinfo {author} {\bibfnamefont {C.}~\bibnamefont
  {Gohle}}, \bibinfo {author} {\bibfnamefont {T.}~\bibnamefont {Udem}},
  \bibinfo {author} {\bibfnamefont {M.}~\bibnamefont {Herrmann}}, \bibinfo
  {author} {\bibfnamefont {J.}~\bibnamefont {Rauschenberger}}, \bibinfo
  {author} {\bibfnamefont {R.}~\bibnamefont {Holzwarth}}, \bibinfo {author}
  {\bibfnamefont {H.~A.}\ \bibnamefont {Schuessler}}, \bibinfo {author}
  {\bibfnamefont {F.}~\bibnamefont {Krausz}}, \ and\ \bibinfo {author}
  {\bibfnamefont {T.~W.}\ \bibnamefont {H{\"a}nsch}},\ }\href@noop {}
  {\bibfield  {journal} {\bibinfo  {journal} {Nature}\ }\textbf {\bibinfo
  {volume} {436}},\ \bibinfo {pages} {234} (\bibinfo {year}
  {2005})}\BibitemShut {NoStop}%
\bibitem [{\citenamefont {Strickland}\ and\ \citenamefont
  {Mourou}(1985)}]{strickland1985compression}%
  \BibitemOpen
  \bibfield  {author} {\bibinfo {author} {\bibfnamefont {D.}~\bibnamefont
  {Strickland}}\ and\ \bibinfo {author} {\bibfnamefont {G.}~\bibnamefont
  {Mourou}},\ }\href@noop {} {\bibfield  {journal} {\bibinfo  {journal} {Opt.
  Commun.}\ }\textbf {\bibinfo {volume} {55}},\ \bibinfo {pages} {447}
  (\bibinfo {year} {1985})}\BibitemShut {NoStop}%
\bibitem [{\citenamefont {Schuller}\ \emph {et~al.}(2010)\citenamefont
  {Schuller}, \citenamefont {Barnard}, \citenamefont {Cai}, \citenamefont
  {Jun}, \citenamefont {White},\ and\ \citenamefont
  {Brongersma}}]{schuller2010plasmonics}%
  \BibitemOpen
  \bibfield  {author} {\bibinfo {author} {\bibfnamefont {J.~A.}\ \bibnamefont
  {Schuller}}, \bibinfo {author} {\bibfnamefont {E.~S.}\ \bibnamefont
  {Barnard}}, \bibinfo {author} {\bibfnamefont {W.}~\bibnamefont {Cai}},
  \bibinfo {author} {\bibfnamefont {Y.~C.}\ \bibnamefont {Jun}}, \bibinfo
  {author} {\bibfnamefont {J.~S.}\ \bibnamefont {White}}, \ and\ \bibinfo
  {author} {\bibfnamefont {M.~L.}\ \bibnamefont {Brongersma}},\ }\href@noop {}
  {\bibfield  {journal} {\bibinfo  {journal} {Nat. Mat.}\ }\textbf {\bibinfo
  {volume} {9}},\ \bibinfo {pages} {193} (\bibinfo {year} {2010})}\BibitemShut
  {NoStop}%
\bibitem [{\citenamefont {Kim}\ \emph {et~al.}(2008)\citenamefont {Kim},
  \citenamefont {Jin}, \citenamefont {Kim}, \citenamefont {Park}, \citenamefont
  {Kim},\ and\ \citenamefont {Kim}}]{kim2008high}%
  \BibitemOpen
  \bibfield  {author} {\bibinfo {author} {\bibfnamefont {S.}~\bibnamefont
  {Kim}}, \bibinfo {author} {\bibfnamefont {J.}~\bibnamefont {Jin}}, \bibinfo
  {author} {\bibfnamefont {Y.-J.}\ \bibnamefont {Kim}}, \bibinfo {author}
  {\bibfnamefont {I.-Y.}\ \bibnamefont {Park}}, \bibinfo {author}
  {\bibfnamefont {Y.}~\bibnamefont {Kim}}, \ and\ \bibinfo {author}
  {\bibfnamefont {S.-W.}\ \bibnamefont {Kim}},\ }\href@noop {} {\bibfield
  {journal} {\bibinfo  {journal} {Nature}\ }\textbf {\bibinfo {volume} {453}},\
  \bibinfo {pages} {757} (\bibinfo {year} {2008})}\BibitemShut {NoStop}%
\bibitem [{\citenamefont {Park}\ \emph {et~al.}(2011)\citenamefont {Park},
  \citenamefont {Kim}, \citenamefont {Choi}, \citenamefont {Lee}, \citenamefont
  {Kim}, \citenamefont {Kling}, \citenamefont {Stockman},\ and\ \citenamefont
  {Kim}}]{park2011plasmonic}%
  \BibitemOpen
  \bibfield  {author} {\bibinfo {author} {\bibfnamefont {I.-Y.}\ \bibnamefont
  {Park}}, \bibinfo {author} {\bibfnamefont {S.}~\bibnamefont {Kim}}, \bibinfo
  {author} {\bibfnamefont {J.}~\bibnamefont {Choi}}, \bibinfo {author}
  {\bibfnamefont {D.-H.}\ \bibnamefont {Lee}}, \bibinfo {author} {\bibfnamefont
  {Y.-J.}\ \bibnamefont {Kim}}, \bibinfo {author} {\bibfnamefont {M.~F.}\
  \bibnamefont {Kling}}, \bibinfo {author} {\bibfnamefont {M.~I.}\ \bibnamefont
  {Stockman}}, \ and\ \bibinfo {author} {\bibfnamefont {S.-W.}\ \bibnamefont
  {Kim}},\ }\href@noop {} {\bibfield  {journal} {\bibinfo  {journal} {Nat.
  Photonics}\ }\textbf {\bibinfo {volume} {5}},\ \bibinfo {pages} {677}
  (\bibinfo {year} {2011})}\BibitemShut {NoStop}%
\bibitem [{\citenamefont {Sivis}\ \emph {et~al.}(2013)\citenamefont {Sivis},
  \citenamefont {Duwe}, \citenamefont {Abel},\ and\ \citenamefont
  {Ropers}}]{sivis2013extreme}%
  \BibitemOpen
  \bibfield  {author} {\bibinfo {author} {\bibfnamefont {M.}~\bibnamefont
  {Sivis}}, \bibinfo {author} {\bibfnamefont {M.}~\bibnamefont {Duwe}},
  \bibinfo {author} {\bibfnamefont {B.}~\bibnamefont {Abel}}, \ and\ \bibinfo
  {author} {\bibfnamefont {C.}~\bibnamefont {Ropers}},\ }\href@noop {}
  {\bibfield  {journal} {\bibinfo  {journal} {Nature Phys.}\ }\textbf {\bibinfo
  {volume} {9}},\ \bibinfo {pages} {304} (\bibinfo {year} {2013})}\BibitemShut
  {NoStop}%
\bibitem [{\citenamefont {Pfullmann}\ \emph {et~al.}(2013)\citenamefont
  {Pfullmann}, \citenamefont {Waltermann}, \citenamefont {Noack}, \citenamefont
  {Rausch}, \citenamefont {Nagy}, \citenamefont {Reinhardt}, \citenamefont
  {Kova\u{c}ev}, \citenamefont {Knittel}, \citenamefont {Bratschitsch},
  \citenamefont {Akemeier}, \citenamefont {H\"utten}, \citenamefont
  {Leitenstorfer},\ and\ \citenamefont {Morgner}}]{pfullmann2013bow}%
  \BibitemOpen
  \bibfield  {author} {\bibinfo {author} {\bibfnamefont {N.}~\bibnamefont
  {Pfullmann}}, \bibinfo {author} {\bibfnamefont {C.}~\bibnamefont
  {Waltermann}}, \bibinfo {author} {\bibfnamefont {M.}~\bibnamefont {Noack}},
  \bibinfo {author} {\bibfnamefont {S.}~\bibnamefont {Rausch}}, \bibinfo
  {author} {\bibfnamefont {T.}~\bibnamefont {Nagy}}, \bibinfo {author}
  {\bibfnamefont {C.}~\bibnamefont {Reinhardt}}, \bibinfo {author}
  {\bibfnamefont {M.}~\bibnamefont {Kova\u{c}ev}}, \bibinfo {author}
  {\bibfnamefont {V.}~\bibnamefont {Knittel}}, \bibinfo {author} {\bibfnamefont
  {R.}~\bibnamefont {Bratschitsch}}, \bibinfo {author} {\bibfnamefont
  {D.}~\bibnamefont {Akemeier}}, \bibinfo {author} {\bibfnamefont
  {A.}~\bibnamefont {H\"utten}}, \bibinfo {author} {\bibfnamefont
  {A.}~\bibnamefont {Leitenstorfer}}, \ and\ \bibinfo {author} {\bibfnamefont
  {U.}~\bibnamefont {Morgner}},\ }\href@noop {} {\bibfield  {journal} {\bibinfo
   {journal} {New J. Phys.}\ }\textbf {\bibinfo {volume} {15}},\ \bibinfo
  {pages} {093027} (\bibinfo {year} {2013})}\BibitemShut {NoStop}%
\bibitem [{\citenamefont {Park}\ \emph {et~al.}(2013)\citenamefont {Park},
  \citenamefont {Choi}, \citenamefont {Lee}, \citenamefont {Han}, \citenamefont
  {Kim},\ and\ \citenamefont {Kim}}]{park2013generation}%
  \BibitemOpen
  \bibfield  {author} {\bibinfo {author} {\bibfnamefont {I.-Y.}\ \bibnamefont
  {Park}}, \bibinfo {author} {\bibfnamefont {J.}~\bibnamefont {Choi}}, \bibinfo
  {author} {\bibfnamefont {D.-H.}\ \bibnamefont {Lee}}, \bibinfo {author}
  {\bibfnamefont {S.}~\bibnamefont {Han}}, \bibinfo {author} {\bibfnamefont
  {S.}~\bibnamefont {Kim}}, \ and\ \bibinfo {author} {\bibfnamefont {S.-W.}\
  \bibnamefont {Kim}},\ }\href@noop {} {\bibfield  {journal} {\bibinfo
  {journal} {Ann. Phys. (Berlin)}\ }\textbf {\bibinfo {volume} {525}},\
  \bibinfo {pages} {87} (\bibinfo {year} {2013})}\BibitemShut {NoStop}%
\bibitem [{\citenamefont {Hutter}\ and\ \citenamefont
  {Fendler}(2004)}]{hutter2004exploitation}%
  \BibitemOpen
  \bibfield  {author} {\bibinfo {author} {\bibfnamefont {E.}~\bibnamefont
  {Hutter}}\ and\ \bibinfo {author} {\bibfnamefont {J.~H.}\ \bibnamefont
  {Fendler}},\ }\href@noop {} {\bibfield  {journal} {\bibinfo  {journal} {Adv.
  Mat.}\ }\textbf {\bibinfo {volume} {16}},\ \bibinfo {pages} {1685} (\bibinfo
  {year} {2004})}\BibitemShut {NoStop}%
\bibitem [{\citenamefont {Keldysh}(1965)}]{keldysh1965ionization}%
  \BibitemOpen
  \bibfield  {author} {\bibinfo {author} {\bibfnamefont {L.}~\bibnamefont
  {Keldysh}},\ }\href@noop {} {\bibfield  {journal} {\bibinfo  {journal} {Sov.
  Phys. JETP}\ }\textbf {\bibinfo {volume} {20}},\ \bibinfo {pages} {1307}
  (\bibinfo {year} {1965})}\BibitemShut {NoStop}%
\bibitem [{\citenamefont {Blaga}\ \emph {et~al.}(2009)\citenamefont {Blaga},
  \citenamefont {Catoire}, \citenamefont {Colosimo}, \citenamefont {Paulus},
  \citenamefont {Muller}, \citenamefont {Agostini},\ and\ \citenamefont
  {DiMauro}}]{blaga2009strong}%
  \BibitemOpen
  \bibfield  {author} {\bibinfo {author} {\bibfnamefont {C.}~\bibnamefont
  {Blaga}}, \bibinfo {author} {\bibfnamefont {F.}~\bibnamefont {Catoire}},
  \bibinfo {author} {\bibfnamefont {P.}~\bibnamefont {Colosimo}}, \bibinfo
  {author} {\bibfnamefont {G.}~\bibnamefont {Paulus}}, \bibinfo {author}
  {\bibfnamefont {H.}~\bibnamefont {Muller}}, \bibinfo {author} {\bibfnamefont
  {P.}~\bibnamefont {Agostini}}, \ and\ \bibinfo {author} {\bibfnamefont
  {L.}~\bibnamefont {DiMauro}},\ }\href@noop {} {\bibfield  {journal} {\bibinfo
   {journal} {Nature Phys.}\ }\textbf {\bibinfo {volume} {5}},\ \bibinfo
  {pages} {335} (\bibinfo {year} {2009})}\BibitemShut {NoStop}%
\bibitem [{\citenamefont {Feti{\'c}}\ \emph {et~al.}(2013)\citenamefont
  {Feti{\'c}}, \citenamefont {Kalajd{\v{z}}i{\'c}},\ and\ \citenamefont
  {Milo{\v{s}}evi{\'c}}}]{fetic2013high}%
  \BibitemOpen
  \bibfield  {author} {\bibinfo {author} {\bibfnamefont {B.}~\bibnamefont
  {Feti{\'c}}}, \bibinfo {author} {\bibfnamefont {K.}~\bibnamefont
  {Kalajd{\v{z}}i{\'c}}}, \ and\ \bibinfo {author} {\bibfnamefont {D.~B.}\
  \bibnamefont {Milo{\v{s}}evi{\'c}}},\ }\href@noop {} {\bibfield  {journal}
  {\bibinfo  {journal} {Ann. Phys. (Berlin)}\ }\textbf {\bibinfo {volume}
  {525}},\ \bibinfo {pages} {107} (\bibinfo {year} {2013})}\BibitemShut
  {NoStop}%
\bibitem [{\citenamefont {Husakou}\ \emph {et~al.}(2011)\citenamefont
  {Husakou}, \citenamefont {Im},\ and\ \citenamefont
  {Herrmann}}]{husakou2011theory}%
  \BibitemOpen
  \bibfield  {author} {\bibinfo {author} {\bibfnamefont {A.}~\bibnamefont
  {Husakou}}, \bibinfo {author} {\bibfnamefont {S.-J.}\ \bibnamefont {Im}}, \
  and\ \bibinfo {author} {\bibfnamefont {J.}~\bibnamefont {Herrmann}},\
  }\href@noop {} {\bibfield  {journal} {\bibinfo  {journal} {Phys. Rev. A}\
  }\textbf {\bibinfo {volume} {83}},\ \bibinfo {pages} {043839} (\bibinfo
  {year} {2011})}\BibitemShut {NoStop}%
\bibitem [{\citenamefont {Ciappina}\ \emph
  {et~al.}(2012{\natexlab{a}})\citenamefont {Ciappina}, \citenamefont
  {Biegert}, \citenamefont {Quidant},\ and\ \citenamefont
  {Lewenstein}}]{ciappina2012high}%
  \BibitemOpen
  \bibfield  {author} {\bibinfo {author} {\bibfnamefont {M.~F.}\ \bibnamefont
  {Ciappina}}, \bibinfo {author} {\bibfnamefont {J.}~\bibnamefont {Biegert}},
  \bibinfo {author} {\bibfnamefont {R.}~\bibnamefont {Quidant}}, \ and\
  \bibinfo {author} {\bibfnamefont {M.}~\bibnamefont {Lewenstein}},\
  }\href@noop {} {\bibfield  {journal} {\bibinfo  {journal} {Phys. Rev. A}\
  }\textbf {\bibinfo {volume} {85}},\ \bibinfo {pages} {033828} (\bibinfo
  {year} {2012}{\natexlab{a}})}\BibitemShut {NoStop}%
\bibitem [{\citenamefont {Yavuz}\ \emph {et~al.}(2012)\citenamefont {Yavuz},
  \citenamefont {Bleda}, \citenamefont {Altun},\ and\ \citenamefont
  {Topcu}}]{yavuz2012generation}%
  \BibitemOpen
  \bibfield  {author} {\bibinfo {author} {\bibfnamefont {I.}~\bibnamefont
  {Yavuz}}, \bibinfo {author} {\bibfnamefont {E.~A.}\ \bibnamefont {Bleda}},
  \bibinfo {author} {\bibfnamefont {Z.}~\bibnamefont {Altun}}, \ and\ \bibinfo
  {author} {\bibfnamefont {T.}~\bibnamefont {Topcu}},\ }\href@noop {}
  {\bibfield  {journal} {\bibinfo  {journal} {Phys. Rev. A}\ }\textbf {\bibinfo
  {volume} {85}},\ \bibinfo {pages} {013416} (\bibinfo {year}
  {2012})}\BibitemShut {NoStop}%
\bibitem [{\citenamefont {P{\'e}rez-Hern{\'a}ndez}\ \emph
  {et~al.}(2013)\citenamefont {P{\'e}rez-Hern{\'a}ndez}, \citenamefont
  {Ciappina}, \citenamefont {Lewenstein}, \citenamefont {Roso},\ and\
  \citenamefont {Za{\"\i}r}}]{perez2013beyond}%
  \BibitemOpen
  \bibfield  {author} {\bibinfo {author} {\bibfnamefont {J.~A.}\ \bibnamefont
  {P{\'e}rez-Hern{\'a}ndez}}, \bibinfo {author} {\bibfnamefont {M.~F.}\
  \bibnamefont {Ciappina}}, \bibinfo {author} {\bibfnamefont {M.}~\bibnamefont
  {Lewenstein}}, \bibinfo {author} {\bibfnamefont {L.}~\bibnamefont {Roso}}, \
  and\ \bibinfo {author} {\bibfnamefont {A.}~\bibnamefont {Za{\"\i}r}},\
  }\href@noop {} {\bibfield  {journal} {\bibinfo  {journal} {Phys. Rev. Lett.}\
  }\textbf {\bibinfo {volume} {110}},\ \bibinfo {pages} {053001} (\bibinfo
  {year} {2013})}\BibitemShut {NoStop}%
\bibitem [{\citenamefont {Ciappina}\ \emph {et~al.}(2013)\citenamefont
  {Ciappina}, \citenamefont {Shaaran},\ and\ \citenamefont
  {Lewenstein}}]{ciappina2013high}%
  \BibitemOpen
  \bibfield  {author} {\bibinfo {author} {\bibfnamefont {M.~F.}\ \bibnamefont
  {Ciappina}}, \bibinfo {author} {\bibfnamefont {T.}~\bibnamefont {Shaaran}}, \
  and\ \bibinfo {author} {\bibfnamefont {M.}~\bibnamefont {Lewenstein}},\
  }\href@noop {} {\bibfield  {journal} {\bibinfo  {journal} {Ann. Phys.
  (Berlin)}\ }\textbf {\bibinfo {volume} {525}},\ \bibinfo {pages} {97}
  (\bibinfo {year} {2013})}\BibitemShut {NoStop}%
\bibitem [{\citenamefont {Yavuz}(2013)}]{yavuz2013gas}%
  \BibitemOpen
  \bibfield  {author} {\bibinfo {author} {\bibfnamefont {I.}~\bibnamefont
  {Yavuz}},\ }\href@noop {} {\bibfield  {journal} {\bibinfo  {journal} {Phys.
  Rev. A}\ }\textbf {\bibinfo {volume} {87}},\ \bibinfo {pages} {053815}
  (\bibinfo {year} {2013})}\BibitemShut {NoStop}%
\bibitem [{\citenamefont {Zhang}\ \emph {et~al.}(2013)\citenamefont {Zhang},
  \citenamefont {Liu},\ and\ \citenamefont {Xu}}]{zhang2013control}%
  \BibitemOpen
  \bibfield  {author} {\bibinfo {author} {\bibfnamefont {C.}~\bibnamefont
  {Zhang}}, \bibinfo {author} {\bibfnamefont {C.}~\bibnamefont {Liu}}, \ and\
  \bibinfo {author} {\bibfnamefont {Z.}~\bibnamefont {Xu}},\ }\href@noop {}
  {\bibfield  {journal} {\bibinfo  {journal} {Phys. Rev. A}\ }\textbf {\bibinfo
  {volume} {88}},\ \bibinfo {pages} {035805} (\bibinfo {year}
  {2013})}\BibitemShut {NoStop}%
\bibitem [{\citenamefont {He}\ \emph {et~al.}(2013)\citenamefont {He},
  \citenamefont {Wang}, \citenamefont {Li}, \citenamefont {Zhang},
  \citenamefont {Lan},\ and\ \citenamefont {Lu}}]{he2013wavelength}%
  \BibitemOpen
  \bibfield  {author} {\bibinfo {author} {\bibfnamefont {L.}~\bibnamefont
  {He}}, \bibinfo {author} {\bibfnamefont {Z.}~\bibnamefont {Wang}}, \bibinfo
  {author} {\bibfnamefont {Y.}~\bibnamefont {Li}}, \bibinfo {author}
  {\bibfnamefont {Q.}~\bibnamefont {Zhang}}, \bibinfo {author} {\bibfnamefont
  {P.}~\bibnamefont {Lan}}, \ and\ \bibinfo {author} {\bibfnamefont
  {P.}~\bibnamefont {Lu}},\ }\href@noop {} {\bibfield  {journal} {\bibinfo
  {journal} {Phys. Rev. A}\ }\textbf {\bibinfo {volume} {88}},\ \bibinfo
  {pages} {053404} (\bibinfo {year} {2013})}\BibitemShut {NoStop}%
\bibitem [{\citenamefont {Luo}\ \emph {et~al.}(2014)\citenamefont {Luo},
  \citenamefont {Li}, \citenamefont {Wang}, \citenamefont {He}, \citenamefont
  {Zhang}, \citenamefont {Lan},\ and\ \citenamefont {Lu}}]{luo2014efficient}%
  \BibitemOpen
  \bibfield  {author} {\bibinfo {author} {\bibfnamefont {J.}~\bibnamefont
  {Luo}}, \bibinfo {author} {\bibfnamefont {Y.}~\bibnamefont {Li}}, \bibinfo
  {author} {\bibfnamefont {Z.}~\bibnamefont {Wang}}, \bibinfo {author}
  {\bibfnamefont {L.}~\bibnamefont {He}}, \bibinfo {author} {\bibfnamefont
  {Q.}~\bibnamefont {Zhang}}, \bibinfo {author} {\bibfnamefont
  {P.}~\bibnamefont {Lan}}, \ and\ \bibinfo {author} {\bibfnamefont
  {P.}~\bibnamefont {Lu}},\ }\href@noop {} {\bibfield  {journal} {\bibinfo
  {journal} {Phys. Rev. A}\ }\textbf {\bibinfo {volume} {89}},\ \bibinfo
  {pages} {023405} (\bibinfo {year} {2014})}\BibitemShut {NoStop}%
\bibitem [{\citenamefont {Ebadi}(2014)}]{ebadi2014interferences}%
  \BibitemOpen
  \bibfield  {author} {\bibinfo {author} {\bibfnamefont {H.}~\bibnamefont
  {Ebadi}},\ }\href@noop {} {\bibfield  {journal} {\bibinfo  {journal} {Phys.
  Rev. A}\ }\textbf {\bibinfo {volume} {89}},\ \bibinfo {pages} {053413}
  (\bibinfo {year} {2014})}\BibitemShut {NoStop}%
\bibitem [{\citenamefont {Lupetti}\ \emph {et~al.}(2013)\citenamefont
  {Lupetti}, \citenamefont {Kling},\ and\ \citenamefont
  {Scrinzi}}]{lupetti2013plasmon}%
  \BibitemOpen
  \bibfield  {author} {\bibinfo {author} {\bibfnamefont {M.}~\bibnamefont
  {Lupetti}}, \bibinfo {author} {\bibfnamefont {M.~F.}\ \bibnamefont {Kling}},
  \ and\ \bibinfo {author} {\bibfnamefont {A.}~\bibnamefont {Scrinzi}},\
  }\href@noop {} {\bibfield  {journal} {\bibinfo  {journal} {Phys. Rev. Lett.}\
  }\textbf {\bibinfo {volume} {110}},\ \bibinfo {pages} {223903} (\bibinfo
  {year} {2013})}\BibitemShut {NoStop}%
\bibitem [{\citenamefont {Stebbings}\ \emph {et~al.}(2011)\citenamefont
  {Stebbings}, \citenamefont {S{\"u}{\ss}mann}, \citenamefont {Yang},
  \citenamefont {Scrinzi}, \citenamefont {Durach}, \citenamefont {Rusina},
  \citenamefont {Stockman},\ and\ \citenamefont
  {Kling}}]{stebbings2011generation}%
  \BibitemOpen
  \bibfield  {author} {\bibinfo {author} {\bibfnamefont {S.~L.}\ \bibnamefont
  {Stebbings}}, \bibinfo {author} {\bibfnamefont {F.}~\bibnamefont
  {S{\"u}{\ss}mann}}, \bibinfo {author} {\bibfnamefont {Y.~Y.}\ \bibnamefont
  {Yang}}, \bibinfo {author} {\bibfnamefont {A.}~\bibnamefont {Scrinzi}},
  \bibinfo {author} {\bibfnamefont {M.}~\bibnamefont {Durach}}, \bibinfo
  {author} {\bibfnamefont {A.}~\bibnamefont {Rusina}}, \bibinfo {author}
  {\bibfnamefont {M.~I.}\ \bibnamefont {Stockman}}, \ and\ \bibinfo {author}
  {\bibfnamefont {M.~F.}\ \bibnamefont {Kling}},\ }\href@noop {} {\bibfield
  {journal} {\bibinfo  {journal} {New J. Phys.}\ }\textbf {\bibinfo {volume}
  {13}},\ \bibinfo {pages} {073010} (\bibinfo {year} {2011})}\BibitemShut
  {NoStop}%
\bibitem [{\citenamefont {Sivis}\ and\ \citenamefont
  {Ropers}(2013)}]{sivis2013generation}%
  \BibitemOpen
  \bibfield  {author} {\bibinfo {author} {\bibfnamefont {M.}~\bibnamefont
  {Sivis}}\ and\ \bibinfo {author} {\bibfnamefont {C.}~\bibnamefont {Ropers}},\
  }\href@noop {} {\bibfield  {journal} {\bibinfo  {journal} {Phys. Rev. Lett.}\
  }\textbf {\bibinfo {volume} {111}},\ \bibinfo {pages} {085001} (\bibinfo
  {year} {2013})}\BibitemShut {NoStop}%
\bibitem [{\citenamefont {Shaaran}\ \emph {et~al.}(2013)\citenamefont
  {Shaaran}, \citenamefont {Ciappina}, \citenamefont {Guichard}, \citenamefont
  {P{\'e}rez-Hern{\'a}ndez}, \citenamefont {Roso}, \citenamefont {Arnold},
  \citenamefont {Siegel}, \citenamefont {Za{\"\i}r},\ and\ \citenamefont
  {Lewenstein}}]{shaaran2013high}%
  \BibitemOpen
  \bibfield  {author} {\bibinfo {author} {\bibfnamefont {T.}~\bibnamefont
  {Shaaran}}, \bibinfo {author} {\bibfnamefont {M.~F.}\ \bibnamefont
  {Ciappina}}, \bibinfo {author} {\bibfnamefont {R.}~\bibnamefont {Guichard}},
  \bibinfo {author} {\bibfnamefont {J.~A.}\ \bibnamefont
  {P{\'e}rez-Hern{\'a}ndez}}, \bibinfo {author} {\bibfnamefont
  {L.}~\bibnamefont {Roso}}, \bibinfo {author} {\bibfnamefont {M.}~\bibnamefont
  {Arnold}}, \bibinfo {author} {\bibfnamefont {T.}~\bibnamefont {Siegel}},
  \bibinfo {author} {\bibfnamefont {A.}~\bibnamefont {Za{\"\i}r}}, \ and\
  \bibinfo {author} {\bibfnamefont {M.}~\bibnamefont {Lewenstein}},\
  }\href@noop {} {\bibfield  {journal} {\bibinfo  {journal} {Phys. Rev. A}\
  }\textbf {\bibinfo {volume} {87}},\ \bibinfo {pages} {041402} (\bibinfo
  {year} {2013})}\BibitemShut {NoStop}%
\bibitem [{\citenamefont {Husakou}\ and\ \citenamefont
  {Herrmann}(2014)}]{husakou2014quasi}%
  \BibitemOpen
  \bibfield  {author} {\bibinfo {author} {\bibfnamefont {A.}~\bibnamefont
  {Husakou}}\ and\ \bibinfo {author} {\bibfnamefont {J.}~\bibnamefont
  {Herrmann}},\ }\href@noop {} {\bibfield  {journal} {\bibinfo  {journal}
  {Phys. Rev. A}\ }\textbf {\bibinfo {volume} {90}},\ \bibinfo {pages} {023831}
  (\bibinfo {year} {2014})}\BibitemShut {NoStop}%
\bibitem [{\citenamefont {Sakabe}\ \emph {et~al.}(2015)\citenamefont {Sakabe},
  \citenamefont {Lienau},\ and\ \citenamefont {Grunwald}}]{sakabe2015progress}%
  \BibitemOpen
  \bibfield  {author} {\bibinfo {author} {\bibfnamefont {S.}~\bibnamefont
  {Sakabe}}, \bibinfo {author} {\bibfnamefont {C.}~\bibnamefont {Lienau}}, \
  and\ \bibinfo {author} {\bibfnamefont {R.}~\bibnamefont {Grunwald}},\
  }\href@noop {} {\emph {\bibinfo {title} {Progress in Nonlinear
  Nano-Optics}}}\ (\bibinfo  {publisher} {Springer International Publishing},\
  \bibinfo {address} {Switzerland},\ \bibinfo {year} {2015})\BibitemShut
  {NoStop}%
\bibitem [{\citenamefont {Shaaran}\ \emph {et~al.}(2012)\citenamefont
  {Shaaran}, \citenamefont {Ciappina},\ and\ \citenamefont
  {Lewenstein}}]{shaaran2012estimating}%
  \BibitemOpen
  \bibfield  {author} {\bibinfo {author} {\bibfnamefont {T.}~\bibnamefont
  {Shaaran}}, \bibinfo {author} {\bibfnamefont {M.~F.}\ \bibnamefont
  {Ciappina}}, \ and\ \bibinfo {author} {\bibfnamefont {M.}~\bibnamefont
  {Lewenstein}},\ }\href@noop {} {\bibfield  {journal} {\bibinfo  {journal} {J.
  Mod. Opt.}\ }\textbf {\bibinfo {volume} {59}},\ \bibinfo {pages} {1634}
  (\bibinfo {year} {2012})}\BibitemShut {NoStop}%
\bibitem [{\citenamefont {Raschke}(2013)}]{raschke2013high}%
  \BibitemOpen
  \bibfield  {author} {\bibinfo {author} {\bibfnamefont {M.~B.}\ \bibnamefont
  {Raschke}},\ }\href@noop {} {\bibfield  {journal} {\bibinfo  {journal} {Ann.
  Phys. (Berlin)}\ }\textbf {\bibinfo {volume} {525}},\ \bibinfo {pages} {A40}
  (\bibinfo {year} {2013})}\BibitemShut {NoStop}%
\bibitem [{\citenamefont {Sivis}\ \emph {et~al.}(2012)\citenamefont {Sivis},
  \citenamefont {Duwe}, \citenamefont {Abel},\ and\ \citenamefont
  {Ropers}}]{sivis2012nanostructure}%
  \BibitemOpen
  \bibfield  {author} {\bibinfo {author} {\bibfnamefont {M.}~\bibnamefont
  {Sivis}}, \bibinfo {author} {\bibfnamefont {M.}~\bibnamefont {Duwe}},
  \bibinfo {author} {\bibfnamefont {B.}~\bibnamefont {Abel}}, \ and\ \bibinfo
  {author} {\bibfnamefont {C.}~\bibnamefont {Ropers}},\ }\href@noop {}
  {\bibfield  {journal} {\bibinfo  {journal} {Nature}\ }\textbf {\bibinfo
  {volume} {485}},\ \bibinfo {pages} {E1} (\bibinfo {year} {2012})}\BibitemShut
  {NoStop}%
\bibitem [{\citenamefont {Watson}\ \emph {et~al.}(1996)\citenamefont {Watson},
  \citenamefont {Sanpera}, \citenamefont {Chen},\ and\ \citenamefont
  {Burnett}}]{Sanpera1}%
  \BibitemOpen
  \bibfield  {author} {\bibinfo {author} {\bibfnamefont {J.}~\bibnamefont
  {Watson}}, \bibinfo {author} {\bibfnamefont {A.}~\bibnamefont {Sanpera}},
  \bibinfo {author} {\bibfnamefont {X.}~\bibnamefont {Chen}}, \ and\ \bibinfo
  {author} {\bibfnamefont {K.}~\bibnamefont {Burnett}},\ }\href@noop {}
  {\bibfield  {journal} {\bibinfo  {journal} {Phys. Rev. A}\ }\textbf {\bibinfo
  {volume} {53}},\ \bibinfo {pages} {R1962} (\bibinfo {year}
  {1996})}\BibitemShut {NoStop}%
\bibitem [{\citenamefont {Sanpera}\ \emph {et~al.}(1996)\citenamefont
  {Sanpera}, \citenamefont {Watson}, \citenamefont {Lewenstein},\ and\
  \citenamefont {Burnett}}]{Sanpera2}%
  \BibitemOpen
  \bibfield  {author} {\bibinfo {author} {\bibfnamefont {A.}~\bibnamefont
  {Sanpera}}, \bibinfo {author} {\bibfnamefont {J.}~\bibnamefont {Watson}},
  \bibinfo {author} {\bibfnamefont {M.}~\bibnamefont {Lewenstein}}, \ and\
  \bibinfo {author} {\bibfnamefont {K.}~\bibnamefont {Burnett}},\ }\href@noop
  {} {\bibfield  {journal} {\bibinfo  {journal} {Phys. Rev. A}\ }\textbf
  {\bibinfo {volume} {54}},\ \bibinfo {pages} {4320} (\bibinfo {year}
  {1996})}\BibitemShut {NoStop}%
\bibitem [{\citenamefont {Moreno}\ \emph {et~al.}(1997)\citenamefont {Moreno},
  \citenamefont {Plaja},\ and\ \citenamefont {Roso}}]{Moreno}%
  \BibitemOpen
  \bibfield  {author} {\bibinfo {author} {\bibfnamefont {P.}~\bibnamefont
  {Moreno}}, \bibinfo {author} {\bibfnamefont {L.}~\bibnamefont {Plaja}}, \
  and\ \bibinfo {author} {\bibfnamefont {L.}~\bibnamefont {Roso}},\ }\href@noop
  {} {\bibfield  {journal} {\bibinfo  {journal} {Phys. Rev. A}\ }\textbf
  {\bibinfo {volume} {55}},\ \bibinfo {pages} {R1593} (\bibinfo {year}
  {1997})}\BibitemShut {NoStop}%
\bibitem [{\citenamefont {Milo\u{s}evi\'c}\ and\ \citenamefont
  {Ehlotzky}(2003)}]{Milosevic}%
  \BibitemOpen
  \bibfield  {author} {\bibinfo {author} {\bibfnamefont {D.}~\bibnamefont
  {Milo\u{s}evi\'c}}\ and\ \bibinfo {author} {\bibfnamefont {F.}~\bibnamefont
  {Ehlotzky}},\ }in\ \href@noop {} {\emph {\bibinfo {booktitle} {Advances in
  Atomic, Molecular and Optical Physics}}},\ \bibinfo {editor} {edited by\
  \bibinfo {editor} {\bibfnamefont {B.}~\bibnamefont {Bederson}}\ and\ \bibinfo
  {editor} {\bibfnamefont {H.}~\bibnamefont {Walther}}}\ (\bibinfo  {publisher}
  {Elsevier Academic Press},\ \bibinfo {address} {Amsterdam},\ \bibinfo {year}
  {2003})\ pp.\ \bibinfo {pages} {373--532}\BibitemShut {NoStop}%
\bibitem [{\citenamefont {Milo\u{s}evi\'c}(2006)}]{Milosevic2}%
  \BibitemOpen
  \bibfield  {author} {\bibinfo {author} {\bibfnamefont {D.}~\bibnamefont
  {Milo\u{s}evi\'c}},\ }\href@noop {} {\bibfield  {journal} {\bibinfo
  {journal} {J. Opt. Soc. Am. B}\ }\textbf {\bibinfo {volume} {23}},\ \bibinfo
  {pages} {308} (\bibinfo {year} {2006})}\BibitemShut {NoStop}%
\bibitem [{\citenamefont {Yuan}\ \emph {et~al.}(2015)\citenamefont {Yuan},
  \citenamefont {Wei}, \citenamefont {Liu}, \citenamefont {Zeng}, \citenamefont
  {Zheng}, \citenamefont {Jiang}, \citenamefont {Ge},\ and\ \citenamefont
  {Li}}]{Yuan}%
  \BibitemOpen
  \bibfield  {author} {\bibinfo {author} {\bibfnamefont {X.}~\bibnamefont
  {Yuan}}, \bibinfo {author} {\bibfnamefont {P.}~\bibnamefont {Wei}}, \bibinfo
  {author} {\bibfnamefont {C.}~\bibnamefont {Liu}}, \bibinfo {author}
  {\bibfnamefont {Z.}~\bibnamefont {Zeng}}, \bibinfo {author} {\bibfnamefont
  {Y.}~\bibnamefont {Zheng}}, \bibinfo {author} {\bibfnamefont
  {J.}~\bibnamefont {Jiang}}, \bibinfo {author} {\bibfnamefont
  {X.}~\bibnamefont {Ge}}, \ and\ \bibinfo {author} {\bibfnamefont
  {R.}~\bibnamefont {Li}},\ }\href@noop {} {\bibfield  {journal} {\bibinfo
  {journal} {Appl. Phys. Lett.}\ }\textbf {\bibinfo {volume} {107}},\ \bibinfo
  {pages} {041110} (\bibinfo {year} {2015})}\BibitemShut {NoStop}%
\bibitem [{\citenamefont {Gallagher}(1994)}]{gallagher2005rydberg}%
  \BibitemOpen
  \bibfield  {author} {\bibinfo {author} {\bibfnamefont {T.~F.}\ \bibnamefont
  {Gallagher}},\ }\href@noop {} {\emph {\bibinfo {title} {Rydberg Atoms}}}\
  (\bibinfo  {publisher} {Cambridge University Press},\ \bibinfo {address}
  {Cambridge},\ \bibinfo {year} {1994})\BibitemShut {NoStop}%
\bibitem [{\citenamefont {Bleda}\ \emph {et~al.}(2013)\citenamefont {Bleda},
  \citenamefont {Yavuz}, \citenamefont {Altun},\ and\ \citenamefont
  {Topcu}}]{bleda2013high}%
  \BibitemOpen
  \bibfield  {author} {\bibinfo {author} {\bibfnamefont {E.~A.}\ \bibnamefont
  {Bleda}}, \bibinfo {author} {\bibfnamefont {I.}~\bibnamefont {Yavuz}},
  \bibinfo {author} {\bibfnamefont {Z.}~\bibnamefont {Altun}}, \ and\ \bibinfo
  {author} {\bibfnamefont {T.}~\bibnamefont {Topcu}},\ }\href@noop {}
  {\bibfield  {journal} {\bibinfo  {journal} {Phys. Rev. A}\ }\textbf {\bibinfo
  {volume} {88}},\ \bibinfo {pages} {043417} (\bibinfo {year}
  {2013})}\BibitemShut {NoStop}%
\bibitem [{\citenamefont {Vitanov}\ \emph {et~al.}(1999)\citenamefont
  {Vitanov}, \citenamefont {Suominen},\ and\ \citenamefont {Shore}}]{Vitanov}%
  \BibitemOpen
  \bibfield  {author} {\bibinfo {author} {\bibfnamefont {N.}~\bibnamefont
  {Vitanov}}, \bibinfo {author} {\bibfnamefont {K.}~\bibnamefont {Suominen}}, \
  and\ \bibinfo {author} {\bibfnamefont {B.}~\bibnamefont {Shore}},\
  }\href@noop {} {\bibfield  {journal} {\bibinfo  {journal} {J. Phys. B}\
  }\textbf {\bibinfo {volume} {32}},\ \bibinfo {pages} {4535} (\bibinfo {year}
  {1999})}\BibitemShut {NoStop}%
\bibitem [{\citenamefont {Vewinger}\ \emph {et~al.}(2003)\citenamefont
  {Vewinger}, \citenamefont {Heinz}, \citenamefont {Fernandez}, \citenamefont
  {Vitanov},\ and\ \citenamefont {Bergmann}}]{Vewinger}%
  \BibitemOpen
  \bibfield  {author} {\bibinfo {author} {\bibfnamefont {F.}~\bibnamefont
  {Vewinger}}, \bibinfo {author} {\bibfnamefont {M.}~\bibnamefont {Heinz}},
  \bibinfo {author} {\bibfnamefont {R.}~\bibnamefont {Fernandez}}, \bibinfo
  {author} {\bibfnamefont {N.}~\bibnamefont {Vitanov}}, \ and\ \bibinfo
  {author} {\bibfnamefont {K.}~\bibnamefont {Bergmann}},\ }\href@noop {}
  {\bibfield  {journal} {\bibinfo  {journal} {Phys. Rev. Lett.}\ }\textbf
  {\bibinfo {volume} {91}},\ \bibinfo {pages} {213001} (\bibinfo {year}
  {2003})}\BibitemShut {NoStop}%
\bibitem [{\citenamefont {Vitanov}\ \emph {et~al.}(2001)\citenamefont
  {Vitanov}, \citenamefont {Halfmann}, \citenamefont {Shore},\ and\
  \citenamefont {Bergmann}}]{Vitanov2}%
  \BibitemOpen
  \bibfield  {author} {\bibinfo {author} {\bibfnamefont {N.}~\bibnamefont
  {Vitanov}}, \bibinfo {author} {\bibfnamefont {T.}~\bibnamefont {Halfmann}},
  \bibinfo {author} {\bibfnamefont {B.}~\bibnamefont {Shore}}, \ and\ \bibinfo
  {author} {\bibfnamefont {K.}~\bibnamefont {Bergmann}},\ }\href@noop {}
  {\bibfield  {journal} {\bibinfo  {journal} {Ann. Rev. Phys. Chem.}\ }\textbf
  {\bibinfo {volume} {52}},\ \bibinfo {pages} {763} (\bibinfo {year}
  {2001})}\BibitemShut {NoStop}%
\bibitem [{\citenamefont {Bergmann}\ \emph {et~al.}(2015)\citenamefont
  {Bergmann}, \citenamefont {Vitanov},\ and\ \citenamefont
  {Shore}}]{Bergmann-review}%
  \BibitemOpen
  \bibfield  {author} {\bibinfo {author} {\bibfnamefont {K.}~\bibnamefont
  {Bergmann}}, \bibinfo {author} {\bibfnamefont {N.~V.}\ \bibnamefont
  {Vitanov}}, \ and\ \bibinfo {author} {\bibfnamefont {B.~W.}\ \bibnamefont
  {Shore}},\ }\href@noop {} {\bibfield  {journal} {\bibinfo  {journal} {J.
  Chem. Phys.}\ }\textbf {\bibinfo {volume} {142}},\ \bibinfo {pages} {170901}
  (\bibinfo {year} {2015})}\BibitemShut {NoStop}%
\bibitem [{\citenamefont {Ciappina}\ \emph
  {et~al.}(2012{\natexlab{b}})\citenamefont {Ciappina}, \citenamefont
  {A\'cimovi\'c}, \citenamefont {Shaaran}, \citenamefont {Biegert},
  \citenamefont {Quidant},\ and\ \citenamefont
  {Lewenstein}}]{ciappina2012optexp}%
  \BibitemOpen
  \bibfield  {author} {\bibinfo {author} {\bibfnamefont {M.~F.}\ \bibnamefont
  {Ciappina}}, \bibinfo {author} {\bibfnamefont {S.~S.}\ \bibnamefont
  {A\'cimovi\'c}}, \bibinfo {author} {\bibfnamefont {T.}~\bibnamefont
  {Shaaran}}, \bibinfo {author} {\bibfnamefont {J.}~\bibnamefont {Biegert}},
  \bibinfo {author} {\bibfnamefont {R.}~\bibnamefont {Quidant}}, \ and\
  \bibinfo {author} {\bibfnamefont {M.}~\bibnamefont {Lewenstein}},\
  }\href@noop {} {\bibfield  {journal} {\bibinfo  {journal} {Opt. Exp.}\
  }\textbf {\bibinfo {volume} {20}},\ \bibinfo {pages} {26261} (\bibinfo {year}
  {2012}{\natexlab{b}})}\BibitemShut {NoStop}%
\bibitem [{\citenamefont {Yavuz}\ \emph {et~al.}(2015)\citenamefont {Yavuz},
  \citenamefont {Tikman},\ and\ \citenamefont {Altun}}]{ilhan2015h2p}%
  \BibitemOpen
  \bibfield  {author} {\bibinfo {author} {\bibfnamefont {I.}~\bibnamefont
  {Yavuz}}, \bibinfo {author} {\bibfnamefont {Y.}~\bibnamefont {Tikman}}, \
  and\ \bibinfo {author} {\bibfnamefont {Z.}~\bibnamefont {Altun}},\
  }\href@noop {} {\bibfield  {journal} {\bibinfo  {journal} {Phys. Rev. A}\
  }\textbf {\bibinfo {volume} {92}},\ \bibinfo {pages} {023413} (\bibinfo
  {year} {2015})}\BibitemShut {NoStop}%
\bibitem [{\citenamefont {Su}\ and\ \citenamefont
  {Eberly}(1991)}]{su1991model}%
  \BibitemOpen
  \bibfield  {author} {\bibinfo {author} {\bibfnamefont {Q.}~\bibnamefont
  {Su}}\ and\ \bibinfo {author} {\bibfnamefont {J.}~\bibnamefont {Eberly}},\
  }\href@noop {} {\bibfield  {journal} {\bibinfo  {journal} {Phys. Rev. A}\
  }\textbf {\bibinfo {volume} {44}},\ \bibinfo {pages} {5997} (\bibinfo {year}
  {1991})}\BibitemShut {NoStop}%
\bibitem [{\citenamefont {Krause}\ \emph {et~al.}(1992)\citenamefont {Krause},
  \citenamefont {Schafer},\ and\ \citenamefont
  {Kulander}}]{krause1992calculation}%
  \BibitemOpen
  \bibfield  {author} {\bibinfo {author} {\bibfnamefont {J.~L.}\ \bibnamefont
  {Krause}}, \bibinfo {author} {\bibfnamefont {K.~J.}\ \bibnamefont {Schafer}},
  \ and\ \bibinfo {author} {\bibfnamefont {K.~C.}\ \bibnamefont {Kulander}},\
  }\href@noop {} {\bibfield  {journal} {\bibinfo  {journal} {Phys. Rev. A}\
  }\textbf {\bibinfo {volume} {45}},\ \bibinfo {pages} {4998} (\bibinfo {year}
  {1992})}\BibitemShut {NoStop}%
\bibitem [{\citenamefont {Burnett}\ \emph {et~al.}(1992)\citenamefont
  {Burnett}, \citenamefont {Reed}, \citenamefont {Cooper},\ and\ \citenamefont
  {Knight}}]{burnett1992calculation}%
  \BibitemOpen
  \bibfield  {author} {\bibinfo {author} {\bibfnamefont {K.}~\bibnamefont
  {Burnett}}, \bibinfo {author} {\bibfnamefont {V.}~\bibnamefont {Reed}},
  \bibinfo {author} {\bibfnamefont {J.}~\bibnamefont {Cooper}}, \ and\ \bibinfo
  {author} {\bibfnamefont {P.}~\bibnamefont {Knight}},\ }\href@noop {}
  {\bibfield  {journal} {\bibinfo  {journal} {Phys. Rev. A}\ }\textbf {\bibinfo
  {volume} {45}},\ \bibinfo {pages} {3347} (\bibinfo {year}
  {1992})}\BibitemShut {NoStop}%
\bibitem [{\citenamefont {Bauer}(1997)}]{bauer1997ejection}%
  \BibitemOpen
  \bibfield  {author} {\bibinfo {author} {\bibfnamefont {D.}~\bibnamefont
  {Bauer}},\ }\href@noop {} {\bibfield  {journal} {\bibinfo  {journal} {Phys.
  Rev. A}\ }\textbf {\bibinfo {volume} {55}},\ \bibinfo {pages} {2180}
  (\bibinfo {year} {1997})}\BibitemShut {NoStop}%
\bibitem [{\citenamefont {Shakeshaft}\ \emph {et~al.}(1990)\citenamefont
  {Shakeshaft}, \citenamefont {Potvliege}, \citenamefont {D{\"o}rr},\ and\
  \citenamefont {Cooke}}]{shakeshaft1990multiphoton}%
  \BibitemOpen
  \bibfield  {author} {\bibinfo {author} {\bibfnamefont {R.}~\bibnamefont
  {Shakeshaft}}, \bibinfo {author} {\bibfnamefont {R.}~\bibnamefont
  {Potvliege}}, \bibinfo {author} {\bibfnamefont {M.}~\bibnamefont {D{\"o}rr}},
  \ and\ \bibinfo {author} {\bibfnamefont {W.}~\bibnamefont {Cooke}},\
  }\href@noop {} {\bibfield  {journal} {\bibinfo  {journal} {Phys. Rev. A}\
  }\textbf {\bibinfo {volume} {42}},\ \bibinfo {pages} {1656} (\bibinfo {year}
  {1990})}\BibitemShut {NoStop}%
\bibitem [{\citenamefont {Barnes}\ \emph {et~al.}(2003)\citenamefont {Barnes},
  \citenamefont {Dereux},\ and\ \citenamefont {Ebbesen}}]{barnes2003surface}%
  \BibitemOpen
  \bibfield  {author} {\bibinfo {author} {\bibfnamefont {W.~L.}\ \bibnamefont
  {Barnes}}, \bibinfo {author} {\bibfnamefont {A.}~\bibnamefont {Dereux}}, \
  and\ \bibinfo {author} {\bibfnamefont {T.~W.}\ \bibnamefont {Ebbesen}},\
  }\href@noop {} {\bibfield  {journal} {\bibinfo  {journal} {Nature}\ }\textbf
  {\bibinfo {volume} {424}},\ \bibinfo {pages} {824} (\bibinfo {year}
  {2003})}\BibitemShut {NoStop}%
\bibitem [{\citenamefont {Eustis}\ and\ \citenamefont
  {El-Sayed}(2006)}]{eustis2006gold}%
  \BibitemOpen
  \bibfield  {author} {\bibinfo {author} {\bibfnamefont {S.}~\bibnamefont
  {Eustis}}\ and\ \bibinfo {author} {\bibfnamefont {M.~A.}\ \bibnamefont
  {El-Sayed}},\ }\href@noop {} {\bibfield  {journal} {\bibinfo  {journal}
  {Chem. Soc. Rev.}\ }\textbf {\bibinfo {volume} {35}},\ \bibinfo {pages} {209}
  (\bibinfo {year} {2006})}\BibitemShut {NoStop}%
\bibitem [{\citenamefont {Willets}\ and\ \citenamefont
  {Van~Duyne}(2007)}]{willets2007localized}%
  \BibitemOpen
  \bibfield  {author} {\bibinfo {author} {\bibfnamefont {K.~A.}\ \bibnamefont
  {Willets}}\ and\ \bibinfo {author} {\bibfnamefont {R.~P.}\ \bibnamefont
  {Van~Duyne}},\ }\href@noop {} {\bibfield  {journal} {\bibinfo  {journal}
  {Annu. Rev. Phys. Chem.}\ }\textbf {\bibinfo {volume} {58}},\ \bibinfo
  {pages} {267} (\bibinfo {year} {2007})}\BibitemShut {NoStop}%
\bibitem [{\citenamefont {Ammosov}\ \emph {et~al.}(1986)\citenamefont
  {Ammosov}, \citenamefont {Delone},\ and\ \citenamefont
  {Kra\u{i}nov}}]{ammosov1986tunnel}%
  \BibitemOpen
  \bibfield  {author} {\bibinfo {author} {\bibfnamefont {M.}~\bibnamefont
  {Ammosov}}, \bibinfo {author} {\bibfnamefont {N.}~\bibnamefont {Delone}}, \
  and\ \bibinfo {author} {\bibfnamefont {V.}~\bibnamefont {Kra\u{i}nov}},\
  }\href@noop {} {\bibfield  {journal} {\bibinfo  {journal} {Sov. Phys. JETP}\
  }\textbf {\bibinfo {volume} {64}},\ \bibinfo {pages} {1191} (\bibinfo {year}
  {1986})}\BibitemShut {NoStop}%
\bibitem [{\citenamefont {Herink}\ \emph {et~al.}(2012)\citenamefont {Herink},
  \citenamefont {Solli}, \citenamefont {Gulde},\ and\ \citenamefont
  {Ropers}}]{herink2012field}%
  \BibitemOpen
  \bibfield  {author} {\bibinfo {author} {\bibfnamefont {G.}~\bibnamefont
  {Herink}}, \bibinfo {author} {\bibfnamefont {D.}~\bibnamefont {Solli}},
  \bibinfo {author} {\bibfnamefont {M.}~\bibnamefont {Gulde}}, \ and\ \bibinfo
  {author} {\bibfnamefont {C.}~\bibnamefont {Ropers}},\ }\href@noop {}
  {\bibfield  {journal} {\bibinfo  {journal} {Nature}\ }\textbf {\bibinfo
  {volume} {483}},\ \bibinfo {pages} {190} (\bibinfo {year}
  {2012})}\BibitemShut {NoStop}%
\bibitem [{\citenamefont {Delone}\ and\ \citenamefont
  {Krainov}(1991)}]{delone1991energy}%
  \BibitemOpen
  \bibfield  {author} {\bibinfo {author} {\bibfnamefont {N.}~\bibnamefont
  {Delone}}\ and\ \bibinfo {author} {\bibfnamefont {V.~P.}\ \bibnamefont
  {Krainov}},\ }\href@noop {} {\bibfield  {journal} {\bibinfo  {journal} {J.
  Opt. Soc. Am. B}\ }\textbf {\bibinfo {volume} {8}},\ \bibinfo {pages} {1207}
  (\bibinfo {year} {1991})}\BibitemShut {NoStop}%
\bibitem [{\citenamefont {Ivanov}\ \emph {et~al.}(1996)\citenamefont {Ivanov},
  \citenamefont {Brabec},\ and\ \citenamefont {Burnett}}]{ivanov1996coulomb}%
  \BibitemOpen
  \bibfield  {author} {\bibinfo {author} {\bibfnamefont {M.~Y.}\ \bibnamefont
  {Ivanov}}, \bibinfo {author} {\bibfnamefont {T.}~\bibnamefont {Brabec}}, \
  and\ \bibinfo {author} {\bibfnamefont {N.}~\bibnamefont {Burnett}},\
  }\href@noop {} {\bibfield  {journal} {\bibinfo  {journal} {Phys. Rev. A}\
  }\textbf {\bibinfo {volume} {54}},\ \bibinfo {pages} {742} (\bibinfo {year}
  {1996})}\BibitemShut {NoStop}%
\bibitem [{\citenamefont {Ciappina}\ \emph {et~al.}(2014)\citenamefont
  {Ciappina}, \citenamefont {P\'erez-Hern\'andez},\ and\ \citenamefont
  {Lewenstein}}]{ciappinacpc}%
  \BibitemOpen
  \bibfield  {author} {\bibinfo {author} {\bibfnamefont {M.~F.}\ \bibnamefont
  {Ciappina}}, \bibinfo {author} {\bibfnamefont {J.~A.}\ \bibnamefont
  {P\'erez-Hern\'andez}}, \ and\ \bibinfo {author} {\bibfnamefont
  {M.}~\bibnamefont {Lewenstein}},\ }\href@noop {} {\bibfield  {journal}
  {\bibinfo  {journal} {Comp. Phys. Comm.}\ }\textbf {\bibinfo {volume}
  {185}},\ \bibinfo {pages} {398} (\bibinfo {year} {2014})}\BibitemShut
  {NoStop}%
\end{thebibliography}
%

\end{document}